\documentclass[prd,aps,onecolumn,preprintnumbers,amsmath,amssymb,superscriptaddress]{revtex4}

\usepackage{amsmath}
\usepackage{graphicx}
\usepackage{dcolumn}
\usepackage{bm}
\usepackage{amssymb}
\usepackage{latexsym}
\usepackage{epstopdf}
\usepackage{threeparttable}
\usepackage{booktabs}
\usepackage{tabularx}
\usepackage{mathrsfs}
\usepackage{hhline}
\usepackage{multirow}
\usepackage{color}
\usepackage[colorlinks,linkcolor=magenta,anchorcolor=blue,citecolor=blue]{hyperref}
\usepackage{ulem}

\def\be{\begin{equation}}
\def\ee{\end{equation}}
\def\bea{\begin{eqnarray}}
\def\eea{\end{eqnarray}}

\begin{document}
\title{Confronting Inflation Models with the Coming Observations on Primordial Gravitational Waves}

\author{Taotao Qiu}
\email{qiutt@mail.ccnu.edu.cn}
\affiliation{Institute of Astrophysics, Central China Normal University, Wuhan 430079, China}
\author{Taishi Katsuragawa}
\email{taishi@mail.ccnu.edu.cn}
\affiliation{Institute of Astrophysics, Central China Normal University, Wuhan 430079, China}
\author{Shulei Ni}
\email{nishulei@mails.ccnu.edu.cn}
\affiliation{Institute of Astrophysics, Central China Normal University, Wuhan 430079, China}
\affiliation{Key Laboratory of Particle Astrophysics, Institute of High Energy Physics, Chinese Academy of Sciences, Beijing 100086, China}

\begin{abstract}
The recent observations from CMB have imposed a very stringent upper-limit on the tensor/scalar ratio $r$ of inflation models, $r < 0.064$, which indicates that the primordial gravitational waves (PGW), even though possible to be detected, should have a power spectrum of a tiny amplitude.
However, current experiments on PGW is ambitious to detect such a signal by improving the accuracy to an even higher level. Whatever their results are, it will give us much information about the early Universe, not only from the astrophysical side but also from the theoretical side, such as model building for the early Universe. 
In this paper, we are interested in analyzing what kind of inflation models can be favored by future observations, 
starting with a kind of general action offered by the effective field theory (EFT) approach. 
We show a general form of $r$ that can be reduced to various models, and more importantly, we show how the accuracy of future observations can put constraints on model parameters by plotting the contours in their parameter spaces.
\end{abstract}

\maketitle
\section{introduction}
The detection of gravitational waves (GWs) has become one of the most important tasks in modern astrophysics and cosmology, not only because it can furtherly examine the correctness of Einstein's General Relativity, 
but also since it can provide an independent probe of our Universe, which is known as the ``standard sirens" \cite{Abbott:2017xzu}. 
Therefore, it is no longer surprising that the first direct detection of GWs from two emerging black holes by the Laser Interferometer Gravitational-Wave Observatory (LIGO) \cite{Abbott:2016blz} has become the hotspot of nowadays science and been awarded the Nobel Prize in Physics in 2017. Since then, LIGO and VIRGO announced several events of black hole GWs \cite{Abbott:2016nmj, Abbott:2017vtc, Abbott:2017gyy, Abbott:2017oio} as well as one event of neutron star GWs \cite{TheLIGOScientific:2017qsa}, which indicate the coming of a new ``gravitational wave era".

However, like the electromagnetic waves, the GWs are distributed over a very wide frequency range, from $10^{-16}$ Hz to $10^{12}$ Hz.
Therefore, to get full information of GWs, we need to utilize various probes for different ranges of frequencies. 
Besides LIGO and VIRGO ($\sim 10^2$Hz), the existing and planning programs for GWs detecting includes FAST ($10^{-8}$Hz$\sim$nHz with annual modulation, same as PTA) \cite{Nan:2011um}, KAGRA (kHz, almost same target-range as those of LIGO and Virgo) \cite{Somiya:2011np}, LISA/TianQin/Taiji (mHz) \cite{AmaroSeoane:2012km, Luo:2015ght, Guo:2018npi}, EPTA (nHz) \cite{Ferdman:2010xq},  AliCPT \cite{Li:2017lat} ($\sim 10^{-16}$Hz) and so on, all of which are devoting themselves on building the ``Multi-band gravitational waves astronomy".

Among the various frequency bands, the most difficult to detect might be the one with ultra-low frequency ($\sim 10^{-16}$Hz) and ultra-long wave-lengths (order of the size of the observational Universe), as well as very low amplitudes. This kind of gravitational waves are believed to be generated at the very early stages of the Universe, probably during the inflationary era \cite{Starobinsky:1979ty, Stewart:1993bc}, and thus dubbed as Primordial Gravitational Waves (PGWs). 
However, these PGWs, also known as primordial tensor perturbations, can affect the CMB photons before the last scattering, and thus, leave hints on the CMB sky map in the form of $B$ mode polarization \cite{Hu:1997hv, Seljak:1996gy, Kamionkowski:1996zd}. Since different evolution of the early Universe can give different evolution behavior of the primordial tensor perturbations and also different features of polarization in the CMB map, the observations of the polarizations$-$therefore, the primordial gravitational waves$-$can be used as a probe to test models for the early Universe, especially, inflation.

Practically, as done in all-sky surveys from WMAP \cite{WMAP} to PLANCK \cite{PLANCK}, the detection on polarizations of CMB photons can be transformed into that on parameters of inflation models, such as the spectral amplitude $A_s$, the spectral index $n_s$, as well as the tensor/scalar ratio $r$. The first two corresponds to the primordial scalar perturbations, while the last one involves both scalar and tensor ones. From the newly-released PLANCK 2018 \cite{Akrami:2018odb}, the constraints on these parameters are $\ln(10^{10}A_s)=3.044\pm0.0014$ ($68\%$ C.L.), $n_s=0.9649\pm0.0042$ ($68\%$ C.L.), (TT, TE, EE+lowE+lensing), $r_{0.002}<0.064$ ($95\%$ C.L., TT, TE, EE+lowE+lensing+BK14). As can be seen from these data, we still can have only upper bound for $r$, which is continuously lowered, although $A_s$ and $n_s$ can be constraints both from above and below. It means that the primordial gravitational waves are really very weak and very difficult to test, and current constraints to primordial gravitational waves, although having been improved much, still needs much more development.

In 2014, we proposed a ground-based CMB experiment called AliCPT in Ali of Tibet, China, which aims to search for PGWs by detecting such $B$ mode polarization \cite{Li:2017lat}. As an experiment in the northern hemisphere, it can cover up to $65\%$ of the sky map, and thus become a very important counterpart to other ground-based experiments, such as that in Chile (Atacama Cosmology Telescope \cite{ACT}, POLARBEAR \cite{POLARBEAR}) and at the South Pole (South Pole Telescope \cite{SPT}, BICEP \cite{BICEP}).

The very ambitious scientific goal of AliCPT is to furtherly improve the sensitivity on $r$-detection and to put a more stringent limit on $r$ by one order of magnitude \cite{Li:2017lat}. The significance of the detection of PGWs will be at least two-folded: If we succeed in detecting PGW, we will have evidence that the PGWs do exist, giving the tensor/scalar ratio be well within the AliCPT region, namely $r\in(0.064,0.01)$. On the other hand, if the PGW is still not detected by then, it means that the upper bound of the tensor/scalar ratio will be lowered again, indicating that inflation models with even smaller tensor/scalar ratio ($r<0.01$) will be favored, examples of which including the Starobinsky model \cite{Starobinsky:1980te}, ultra-slow-roll inflation model \cite{Kinney:2005vj}/constant-roll inflation model \cite{Motohashi:2014ppa}, among many other models in the literature.

In this paper, we try to investigate that for various inflation models, how the tensor/scalar ratio can be made small, especially, to meet with the forthcoming observational data. 
In order to do this, we start with a general form of action, offered by the effective field theory (EFT) approach \cite{Cheung:2007st, Gubitosi:2012hu, Gleyzes:2013ooa, Piazza:2013coa, Kase:2014cwa, Gao:2014soa, Cai:2016thi, Cai:2017tku}. 
It has been proved in \cite{Gleyzes:2013ooa, Kase:2014cwa, Cai:2016thi, Cai:2017tku} that the EFT action they are using is very general at least up to the quadratic level, and be able to cover a large class of field actions with the second-order equation of motion, such as Horndeski \cite{Horndeski:1974wa} and GLPV theories \cite{Gleyzes:2014dya}. 
Therefore, we can obtain a general form of the tensor/scalar ratio $r$ that can be reduced to various concrete models, and this allows us to discuss constraints on parameters of those models.
Note that similar work has been done in \cite{DeFelice:2014bma} with more focus on the tensor perturbation itself, such as power spectrum and spectral index.

The rest of the paper is organized as the following: in Sec. II, we introduce the general action in light of the work in \cite{Gleyzes:2013ooa, Kase:2014cwa, Cai:2016thi, Cai:2017tku}, 
and by calculating both scalar and tensor perturbations, we find a very general expression of the tensor/scalar ratio.
In Sec. III, we apply our results by reducing the general form of $r$ to concrete examples. 
We show relations with various slow-varying parameters for each model and obtain the range of each pair of parameters, requiring $r$ be within regions of detection/non-detection of PGWs. We also show how $r$ of these models can deviate from the usual consistency relation in inflation models. 
Sec. V includes our final remarks and discussions.

\section{From The General Inflation Action to the Tensor/Scalar Ratio}
Based on the metric in the ADM form:
\be
\label{metric}
ds^2=-N^2dt^2+h_{ij}(dx^i+N^idt)(dx^j+N^jdt)~
\ee
with $N$ and $N^i$ the lapse function and shift vector while $h_{ij}$ the 3-dimentional spatial metric, 
the very general action up to quadratic perturbation level (in EFT form) is given by \cite{Cheung:2007st, Gleyzes:2013ooa, Kase:2014cwa, Cai:2016thi, Cai:2017tku}:
\bea
\label{eft_action}
S&=&\int d^4x\sqrt{-g}\Big[{M_p^2\over2} f(t)R-\Lambda(t)-c(t)g^{00}~\nonumber\\
&&+{m_2^4(t)\over2}(\delta g^{00})^2-{m_3^3(t)\over2}\delta
K\delta g^{00} -m_4^2(t)\left( \delta K^2-\delta K_{\mu\nu}\delta
K^{\mu\nu} \right) +{\tilde{m}_4^2(t)\over 2}R^{(3)}\delta g^{00}~\nonumber\\
&&-\bar{m}_4^2(t)\delta K^2+{\bar{m}_5(t)\over 2}R^{(3)}\delta K
+{\bar{\lambda}(t)\over2}(R^{(3)})^2+...~\nonumber\\
&&-{\tilde{\lambda}(t)\over M_p^2}\nabla_iR^{(3)}\nabla^iR^{(3)} +... \Big]~,
\eea
The first line is background and the rest are for the perturbations up to second order. Note that according to the merit of EFT approach, the action is clearly written order by order of the perturbations, and the ellipse denotes all orders higher than 2. In the action, we define $\delta K_{\mu\nu}=K_{\mu\nu}-H \Theta_{\mu\nu},~\delta K=K-3H$, where the induced metric $\Theta_{\mu\nu}\equiv g_{\mu\nu}+n_\mu n_\nu$ and the normal vector is defined as $n_\mu\equiv(-N,0,0,0)$. Moreover, since the third and the fourth lines are for higher space (but not time) derivatives, in the following analysis we turned them off by setting $\bar{m}_4=\bar{m}_5=\bar{\lambda}=\tilde{\lambda}=0$.

\subsection{the background equations of motion}
The background of the metric (\ref{metric}) is of the well-known FLRW form, which is in the diagonal form of $\{-1,a^2(t), a^2(t), a^2(t)\}$. It is straightforward to get the background equations from action (\ref{eft_action}), by varying the first line with respect to the lapse function $N$ and the scale factor $a$:
\bea
\label{friedmannbg1}
3M_p^2[f(t)H^2+\dot f(t)H]&=&c(t)+\Lambda(t)~,\\
\label{friedmannbg2}
-M_p^2[2f(t)\dot H+3f(t)H^2+2\dot f(t)H+\ddot
f(t)]&=&c(t)-\Lambda(t)~.
\eea
These are actually nothing but the Friedmann equations, and for the minimal coupling theories where $f(t)=1$, one can have $c(t)=-M_p^2{\dot H}$ and $\Lambda(t)=M_p^2({\dot H}+3H^2)$, which are the same as the results obtained in \cite{Cheung:2007st}. 
In the nontrivial case where $f(t)$ is an arbitrary function, the theory is extended to include also cases where gravity part is modified, or there is nonminimally coupling between the field and gravity parts. From Eq.s (\ref{friedmannbg1}) and (\ref{friedmannbg2}) one can get:
\bea
\label{Hubble}
H(t)&=&-\frac{\dot f}{2f}\pm \frac{\sqrt{3}}{6}\sqrt{3\left(\frac{\dot f}{f}\right)^2+4\frac{(c+\Lambda)}{M_p^2f}}~,\\
\label{Hubbledot}
\dot H(t)&=&-\frac{c}{M_p^2f}-\frac{\ddot f}{2f}-\left(\frac{\dot f}{2f}\right)^2\pm\frac{\sqrt{3}}{12}\sqrt{3\left(\frac{\dot f}{f}\right)^4+4\left(\frac{\dot f}{f}\right)^2\frac{(c+\Lambda)}{M_p^2f}}~\nonumber\\
&=&-\frac{c}{M_p^2f}-\frac{\ddot f}{2f}+\frac{H\dot f}{2f}~.\eea

Defining $\delta_{f}^{(1)}\equiv\dot{f}/Hf$, $\delta_{f}^{(2)}\equiv\ddot{f}/H\dot{f}$, one can also get a neat form of the Hubble parameter squared from Eq. (\ref{Hubble}) as:
\be
H^2=\frac{c+\Lambda}{3M_p^2f}+\frac{1}{2}H^2(\delta_{f}^{(1)})^2\mp\frac{\sqrt{3}}{6}H^2\sqrt{3(\delta_{f}^{(1)})^4+4(\delta_{f}^{(1)})^2\frac{(c+\Lambda)}{H^2M_p^2f}}~,
\ee
which gives the solution
\be
\frac{c+\Lambda}{M_p^2H^2f}=3(1+\delta_{f}^{(1)})~.
\ee
Moreover, the slow-roll parameter can be written as:
\bea
\label{epsilon}
\epsilon\equiv-\frac{\dot H}{H^2}=\frac{3c}{c+\Lambda}(1+\delta_f^{(1)})+\frac{1}{2}\delta_{f}^{(1)}(\delta_{f}^{(2)}-1)~.
\eea
which will be frequently used in the following analysis.

\subsection{scalar perturbation}
Using the action (\ref{eft_action}) and taking the unitary gauge, we find the quadratic action of the scalar perturbation \cite{Cai:2016thi}:
\be
\label{scalar}
S^{(2)}_\zeta=\int d^4x a^3\left[c_1\dot\zeta^2-\left(\frac{\dot c_3}{a}-c_2\right)\frac{(\partial\zeta)^2}{a^2}\right]~,
\ee
where $\zeta$ is the curvature perturbation coming from the scalar perturbation in metric (\ref{metric}), and
\bea
\label{c1}
c_1&=&
\frac{1}{D}\left(2m_4^2+fM_p^2\right)\Big\{3m_3^6+4f^2H^2\epsilon M_p^4+16m_ 2^4m_4^2~\nonumber\\
&&+M_p^2\left[-4\ddot fm_4^2+\dot f\left(-6m_3^3+4Hm_4^2+3\dot fM_p^2\right)\right]~\nonumber\\
&&+2fM_p^2\left[4m_2^4-\ddot fM_p^2+H\left(4H\epsilon m_4^2+\dot fM_p^2\right)\right]\Big\}~,\\
\label{c2}
c_2&=&fM_p^2~,\\
\label{c3}
c_3&=&\frac{2a}{D}\left(2m_4^2+fM_p^2\right)\Big\{2f^2HM_p^4+fM_p^2\left[-m_3^3+\dot fM_p^2+4Hm_4^2\right]\Big\}~,\\
\label{c4}
D&=&\left[m_ 3^3-4Hm_4^2-\left(2fH+\dot f\right)M_p^2\right]^2~.
\eea
According to action (\ref{scalar}), one can get the equation of motion:
\be
\label{eomscalar}
u^{\prime\prime}+c_{s}^{2}k^{2}u-\frac{z^{\prime\prime}}{z}u=0~,
\ee
where $u\equiv z\zeta$, $z\equiv a\sqrt{c_1}$, and prime denotes derivative with respect to the conformal time: $\eta\equiv\int a^{-1}dt$. 
The sound speed squared is also defined as:
\be
c_s^2\equiv\left(\frac{\dot c_3}{a}-c_2\right)\Big/c_1~.
\ee

For initial condition, we consider the case of the subhorizon region $c_s^2k^2\gg z^{\prime\prime}/z$, and we assume that the adiabatic condition $|\omega^\prime/\omega^2|\ll 1$ is satisfied, where $\omega^2\equiv c_s^2k^2-z^{\prime\prime}/z$, which is true for wide range of parameter choice. Therefore, one can apply the WKB approximation to get:
\be
\label{initial}
u_{ini}=\frac{1}{\sqrt{2c_sk}}e^{i\int c_skd\eta}~.
\ee
On the other hand, for a whole solution, assuming $z\propto\eta^{\frac{1}{2}-\nu}$, where the parameter $\nu$ is assumed to be a constant. Moreover, for simplicity but without losing generality, we assume the sound speed squared $c_s\propto \eta^s$, then Eq. (\ref{eomscalar}) becomes
\be
u^{\prime\prime}+c_{s}^{2}k^{2}u-\frac{4\nu^{2}-1}{4\eta^{2}}u=0~,
\ee
and the solution is the famous Hankel function:
\be
u=C\sqrt{\eta}\left[H_{\nu/(s+1)}^{(1)}\left(\Big|\int c_{s}kd\eta\Big|\right)+H_{-\nu/(s+1)}^{(1)}\left(\Big|\int c_{s}kd\eta\Big|\right)\right]~,
\ee
with $C=\sqrt{\pi/(s+1)}/2$ comparing to the initial condition (\ref{initial}). 
In the superhorizon region $c_s^2k^2\ll z^{\prime\prime}/z$, one has $H_{\nu}(\int c_skd\eta)=\sqrt{2/\pi}(\int c_skd\eta)^{-\nu}$, therefore we have
\bea
u&=&\sqrt{\frac{\eta}{2(s+1)}}\left[\left(\int c_{s}kd\eta\right)^{-\frac{\nu}{s+1}}+\left(\int c_{s}kd\eta\right)^{\frac{\nu}{s+1}}\right]~,\nonumber\\
\zeta=\frac{u}{z}&\propto&\frac{\eta^{\nu}}{\sqrt{2(s+1)}}\left(\int c_{s}kd\eta\right)^{-\frac{\nu}{s+1}}\left[1+\left(\int c_{s}kd\eta\right)^{\frac{2\nu}{s+1}}\right]~,
\eea
and the power spectrum is
\bea
P_{\zeta}&\equiv&\frac{k^{3}}{2\pi^{2}}\Big|\frac{u}{z}\Big|^{2}~\nonumber\\
&=&\frac{k^{3}}{2\pi^{2}}\frac{\eta}{2(s+1)a^{2}c_1}\left[\left(\int c_{s}kd\eta\right)^{-\frac{\nu}{s+1}}+\left(\int c_{s}kd\eta\right)^{\frac{\nu}{s+1}}\right]^{2}~.
\eea

We assume slow-varying variable $\epsilon\equiv-\dot H/H^2$, therefore it is easy to get $a\sim\eta^{1/(\epsilon-1)}$, and also $(aH)^{-1}=(\epsilon-1)\eta$. Moreover, we set $c_s=c_{s\ast}(\eta/\eta_\ast)^s$ where $\ast$ denotes some normalization scale, therefore
\be
\label{scalarspectrum}
P_{\zeta}=\frac{(s+1)^2(\epsilon-1)^{2}H_{\ast}^{2}}{4\pi^{2}c_{1\ast}c_{s\ast}^{3}}\left(\frac{\eta}{\eta_{\ast}}\right)^{-3+2\nu-3s}\left(\frac{c_{s}k\eta}{s+1}\right)^{3-\frac{2\nu}{s+1}}\left[1+\left(\frac{c_{s}k\eta}{s+1}\right)^{\frac{2\nu}{s+1}}\right]^{2}
\ee

The current observations indicated that the power spectrum of the scalar perturbation (\ref{scalarspectrum}) should be (nearly) scale-invariant. In order to be so, one can either have $\nu/(s+1)\simeq3/2$, with $\left(\frac{c_{s}k\eta}{s+1}\right)^{\frac{2\nu}{s+1}}$ decreasing:
\be
\label{scalarconstant}
P_{\zeta}=\frac{(s+1)^2(\epsilon-1)^{2}H_{\ast}^{2}}{4\pi^{2}c_{1\ast}c_{s\ast}^{3}}~,
\ee
which is still time-invariant, or have $\nu/(s+1)\simeq-3/2$, with $\left(\frac{c_{s}k\eta}{s+1}\right)^{\frac{2\nu}{s+1}}$ increasing:
\be
\label{scalargrowing}
P_{\zeta}=\frac{(s+1)^2(\epsilon-1)^{2}H_{\ast}^{2}}{4\pi^{2}c_{1\ast}c_{s\ast}^{3}}\left(\frac{\eta}{\eta_{\ast}}\right)^{-6(s+1)}~,
\ee
which will be proportional to $\eta^{-6}$ for constant $c_s$. 

\subsection{tensor perturbation}
We can perform the same procedure to get the power spectrum for tensor perturbations. According to action (\ref{eft_action}), one  can also obtain the quadratic action of the tensor perturbation \cite{Cai:2016thi}:
\be
\label{tensor}
S^{(2)}_T=\frac{M_p^2}{8}\int d^4x a^3{\cal D}_T\left[\dot\gamma_{ij}^2-c_T^2\frac{(\partial_k\gamma_{ij})^2}{a^2}\right]~,
\ee
where $\gamma_{ij}$ is the tensor perturbation in metric (\ref{metric}), and
\be
{\cal D}_T=f+2\frac{m_4^2}{M_p^2}~,~~~c_T^2=\frac{f}{{\cal D}_T}~.
\ee

One can also get the equation of motion:
\be
v^{\prime\prime}+c_T^2k^2v-\frac{z_T^{\prime\prime}}{z_T}=0~,
\ee
where $v\equiv z_T\gamma_{+,\times}$ where $\gamma_{+,\times}$ are two polarization modes of $\gamma_{ij}$, $z_T^2\equiv a^2{\cal D}_T$. Following the same procedure as of the scalar perturbation, and assuming $z_T\propto\eta^{\frac{1}{2}-\nu_T}$, one gets the solution:
\bea
v&=&\sqrt{\frac{\eta}{2(s_T+1)}}\left[\left(\int c_{T}kd\eta\right)^{-\frac{\nu_T}{s_T+1}}+\left(\int c_{T}kd\eta\right)^{\frac{\nu_T}{s_T+1}}\right]~,\nonumber\\
\gamma=\frac{v}{z_T}&\propto&\frac{\eta^{\nu_T}}{\sqrt{2(s_T+1)}}\left(\int c_{T}kd\eta\right)^{-\frac{\nu_T}{s_T+1}}\left[1+\left(\int c_{T}kd\eta\right)^{\frac{2\nu_T}{s_T+1}}\right]~,
\eea
where we assume $c_T=c_{T\ast}(\eta/\eta_\ast)^{s_T}$, and the power spectrum is:
\bea
\label{tensorspectrum}
P_{T}&\equiv&2\frac{k^{3}}{2\pi^{2}}\Big|\frac{v}{z_{T}}\Big|^{2}\nonumber\\
&=&\frac{(s_T+1)^2(\epsilon-1)^{2}H^{2}}{2\pi^{2}{\cal D}_{T}c_{T}^{3}}\left(\frac{\eta}{\eta_{\ast}}\right)^{-3+2\nu_T-3s_T}\left(\frac{c_{T}k\eta}{s_T+1}\right)^{3-\frac{2\nu_{T}}{s_T+1}}\left[1+\left(\frac{c_{T}k\eta}{s_T+1}\right)^{\frac{2\nu_{T}}{s_T+1}}\right]^{2}~.
\eea

The current observations have not provided constraint on the scale variance of primordial tensor power spectrum yet. 
However, in this work, we restrict ourselves on the case where the tensor spectrum is also scale-invariant, as is for the scalar one. In order to be so, one can either have $\nu_T/(s_T+1)\simeq 3/2$, with $\left(\frac{c_{T}k\eta}{s_T+1}\right)^{\frac{2\nu_{T}}{s_T+1}}$ decreasing:
\be
\label{tensorconstant}
P_{T}=\frac{(s_{T}+1)^{2}(\epsilon-1)^{2}H_{\ast}^{2}}{2\pi^{2}\mathcal{D}_{T\ast}c_{T\ast}^{3}}
\ee
which is still time-invariant, or have $\nu_T/(s_T+1)\simeq-3/2$, with $\left(\frac{c_{T}k\eta}{s_T+1}\right)^{\frac{2\nu_{T}}{s_T+1}}$ increasing:
\be
\label{tensorgrowing}
P_{T}=\frac{(s_{T}+1)^{2}(\epsilon-1)^{2}H_{\ast}^{2}}{2\pi^{2}\mathcal{D}_{T\ast}c_{T\ast}^{3}}\left(\frac{\eta}{\eta_{\ast}}\right)^{-6(s_{T}+1)}
\ee
which will be proportional to $\eta^{-6}$ for constant $c_T$. 
Moreover, it is easy to see that, for $\nu_T/(s_T+1)<-3/2$ or $\nu_T/(s_T+1)>3/2$, the spectrum would have a red tilt, while for $-3/2<\nu_T/(s_T+1)<3/2$, the spectrum would have a blue tilt.

\subsection{tensor/scalar ratio}
The tensor/scalar ratio is defined as:
\be
r\equiv \frac{P_T}{P_\zeta}~,
\ee
where in general, $P_T$ and $P_\zeta$ is given in Eqs. (\ref{scalarspectrum}) and (\ref{tensorspectrum}). Hereafter, for simplicity, we stick ourselves only on the cases where tensor spectrum is also scale-invariant. According to the above analysis, one can immediately get the tensor/scalar ratio:
\be
r=2\frac{\mathcal{D}_{s}c_{s}^{3}(s_{T}+1)^{2}}{\mathcal{D}_{T}c_{T}^{3}(s+1)^{2}}~.
\ee

Moreover, since
\be
{\cal D}_{s}=c_1~,~
c_s=\frac{\sqrt{\dot c_3/a-c_2}}{\sqrt{\cal D}_{s}}~,~
{\cal D}_{T}=\frac{M_p^2}{8}\Bigg[f+2\frac{m_4^2}{M_p^2}\Bigg]~,~
c_{T}=\frac{M_p\sqrt{f}}{\sqrt{8{\cal D}_T}}~,
\ee
one can express $r$ in a very general form, namely
\be
\label{r_final}
r=16\frac{(\dot c_3/a-c_2)^{3/2}\sqrt{f+2(m_4/M_p)^2}(s_T+1)^2}{\sqrt{c_1}f^{3/2}M_p^2(s+1)^2}~.
\ee
Note that the above is our master formula on $r$. 
At this stage, it contains EFT functions only, and by reducing them to field functions of concrete models, as we will do below, it can be directly related to the slow-varying parameters of each model. 
The constraints on those models, especially on those parameters, can thus be obtained, via the future constraint on $r$.

Before heading to the next section, let us make further comments on Eq. (\ref{r_final}): although the relation between $r$ and various parameters seems obscure, it is clear with two parameters, $s_T$ and $s$, which represents the time-dependence of the sound speed with respect to tensor and scalar perturbations, by the previous definition. Marginalizing the effects of other parameters, it is evident that a large running of $c_T$ will make $r$ large, while that of $c_s$ will do the opposite. Due to the above, in the following, we will ignore these two parameters for simplicity, by assuming that both $c_T$ and $c_s$ are slow-varying. This is a very common-used assumption in the analysis of inflation models.

\section{Confronting $r$ with future constraints: concrete examples}

\subsection{Towards the concreteness: An dictionary}

In this section, we will apply Eq. (\ref{r_final}) to concrete models, trying to discuss how $r$ is related to usually defined slow-roll parameters, and what the parameters will be like for $r$ to be within these regions. 
Written in the form of Eq.~\eqref{r_final}, the upper limit on the tensor/scalar ratio $r$ provides us with the constraints on the functions $f(t)$, $\Lambda(t)$, $c(t)$, $m_{i}(t)$, and $\tilde{m}_{4}(t)$. 
Although we can get allowed region in multidimensional parameter space, 
there are so many degrees of freedom, and it is rather difficult to handle it in an analytic way. 
Moreover, for different models, the constrained parameter spaces will be different due to different marginalization methods, 
so it becomes meaningless to constrain those general functions directly. 
In order to be specific and reduce the degrees of freedom, it is useful to consider the constraints after reducing to concrete models. 

On the other hand, we do have some tools for such a reduction. For instance, there is a large group of field models which possess the advantage of being ghost-free, by making their equation of motion to be at most second order. Most of these models can be summarized into the Generalized Scalar-Tensor (GST, a.k.a. Horndeski) theory, namely \cite{Deffayet:2009mn}
\bea
\label{GST2}
{\cal L}_2&=&G_2(\phi, X)~,\\
\label{GST3}
{\cal L}_3&=&G_{3}(\phi,X)\Box\phi~,\\
\label{GST4}
{\cal L}_4&=&G_{4}(\phi,X)R-2G_{4X}[(\Box\phi)^{2}-\phi^{;\mu\nu}\phi_{;\mu\nu}]~,\\
\label{GST5}
{\cal L}_5&=&G_{5}(\phi,X)G_{\mu\nu}\phi^{;\mu\nu}+\frac{1}{3}G_{5X}(\phi,X)[(\Box\phi)^{3}-3\Box\phi\phi_{;\mu\nu}\phi^{;\mu\nu}+2\phi_{;\mu\nu}\phi^{;\mu\sigma}\phi_{\ ;\sigma}^{;\mu}]~
\eea
with $X=-1/2 (\partial_{\mu} \phi)^{2}$ and ``;" denotes covariant derivative. Albeit somehow general, it will be rather tedious to calculate the observables using this form, let alone more and more generalized forms are still being developed. However, as has been shown in \cite{Gleyzes:2013ooa, Kase:2014cwa}, one can create a dictionary between our generalized form and GST theories, by using which,  the concrete models can be directly ``read off". Therefore, one can immediately get the expression for various models, without bothering to calculate them one by one. 

For the GST theory whose Lagrangian is given by Eqs. (\ref{GST2})-(\ref{GST5}), the dictionary turns out to be:
\bea
\label{lambda}
\Lambda&=&M_p^2(6fH^2+5\dot{f}H+2f\dot{H}+\ddot{f})/2~,\\
\label{c}
c&=&M_p^2(\dot{f}H-2f\dot{H}-\ddot{f})/2~,\\
\label{m2}
m_2^4&=&(2X)^{3/2}[\sqrt{2X}(E_2+3HE_3+12H^2XE_{4,X}-6H^2E_4-2H^3XE_{5,X})_{,X}]_{,X}/4-c/2~,\\
\label{m3}
m_3^3&=&\dot{f}+2XE_{3,X}+8HX(2XE_{4,X}-E_4)_{,X}-4H^2X(XE_{5,X})_{,X}~,\\
\label{m4}
m_4^2&=&\tilde{m}_4^2=-2XE_{4,X}+HXE_{5,X}-\dot{E}_5/2~,\\
\label{f}
M_p^2f&=&2E_4+\dot{E}_5~,
\eea 
where 
\bea
&&E_2=G_2+\sqrt{X}\int\frac{G_{3,\phi}}{\sqrt{X}}dX~,~E_3=-\int\sqrt{2X}G_{3,X}dX-2\sqrt{2X}G_{4,\phi}~,\nonumber\\
&&E_4=G_4-\sqrt{X}\int\frac{G_{5,\phi}}{2\sqrt{X}}dX~,~E_5=-\int\sqrt{2X}G_{5,X}dX~.
\eea
This dictionary coincides with \cite{Gleyzes:2013ooa, Kase:2014cwa} at least on quadratic perturbation level. From this we can express Eq. (\ref{r_final}) for various concrete models, as will be shown in the following. Moreover, the dictionary can also be applied to models beyond (\ref{GST2})-(\ref{GST5}). For example, for the GLPV model where the Lagrangian is enlarged as \cite{Gleyzes:2014dya}
\be
{\cal L}_4\rightarrow {\cal L}_4+F_4(\phi,X)\epsilon^{\mu\nu\rho}_{~~~\sigma}\epsilon^{\mu^\prime\nu^\prime\rho^\prime\sigma}\phi_{;\mu}\phi_{;\mu^\prime}\phi_{;\nu\nu^\prime}\phi_{;\rho\rho^\prime}~,~{\cal L}_5\rightarrow {\cal L}_5+F_5(\phi,X)\epsilon^{\mu\nu\rho\sigma}\epsilon^{\mu^\prime\nu^\prime\rho^\prime\sigma}\phi_{;\mu}\phi_{;\mu^\prime}\phi_{;\nu\nu^\prime}\phi_{;\rho\rho^\prime}\phi_{;\sigma\sigma^\prime}~
\ee
Eq. (\ref{m2}), (\ref{m3}) and (\ref{m4}) in the dictionary will be enlarged to 
\bea
m_2^4&=&(2X)^{3/2}[\sqrt{2X}(E_2+3HE_3+12H^2XE_{4,X}-6H^2E_4-24H^2X^2F_4\nonumber\\
&&-2H^3XE_{5,X}+6H^3(2X)^{5/2}F_5)_{,X}]_{,X}/4-c/2~,\\
m_3^3&=&\dot{f}+2XE_{3,X}+8HX(2XE_{4,X}-E_4-4X^2F_4)_{,X}-4H^2X(XE_{5,X}+3(2X)^{5/2}F_5)_{,X}~,\\
m_4^2&=&-2XE_{4,X}+HXE_{5,X}+4X^2F_4-3H(2X)^{5/2}F_5-\dot{E}_5/2~,\\
\tilde{m}_4^2&=&-2XE_{4,X}+HXE_{5,X}-\dot{E}_5/2~,
\eea

As a remark, one may notice that for Horndeski theories, we always have $m_4^2=\tilde{m}_4^2$, however for theories beyond, we may not. 
Nonetheless, one may also see that, for the additional function satisfying certain conditions, such as $F_4=3H\sqrt{2X}F_5$, we still have $m_4^2=\tilde{m}_4^2$, that is to say, action (\ref{eft_action}) with $m_4^2=\tilde{m}_4^2$ can not only cover Horndeski theory, but also part of GLPV theory (but not all). It is interesting because, apparently, adding the condition will make the number of functions in action (\ref{eft_action}) (five, because the function $\Lambda$ is totally of background and doesn't enter into the perturbation level) smaller than those in field theory (GLPV) (six). 
On the other hand, if one releases such a condition, 
action (\ref{eft_action}) can fully cover GLPV theory since the number of functions of both sides is the same (six each).  

For the beyond Horndeski models which will cause difference between $m_4^2$ and $\tilde{m}_4^2$, for simplicity and without losing generality, we use a unified form to describe the beyond part, namely \cite{Cai:2016thi, Cai:2017tku}
\be
\label{GST6}
{\cal L}_6=\Delta\tilde{m}_{4}^{2}R^{(3)}\delta g^{00}~,
\ee
where $\Delta\tilde{m}_{4}^{2}=\tilde{m}_4^2-m_4^2$.
In the following, we will choose models which contain two or three of those above Lagrangian, as simple examples. 
We will study how the concrete model Lagrangians (\ref{GST2}), (\ref{GST3}),(\ref{GST4}), (\ref{GST5}), (\ref{GST6}) are written in terms of the EFT functions systematically, 
and we will discuss the constraints on model parameters and plot the allowed regions.

\subsection{K-essence: $M_{p}^{2}R/2+L_{2}$}
We first start with the simplest case of K-essence single field \cite{ArmendarizPicon:1999rj}. 
This case can be written in terms of action (\ref{eft_action}) with the correspondence:
\be
f=1~,~\Lambda=\Lambda_{(2)}=\frac{\dot{\phi}^{2}}{2}G_{2X}(\phi,X)-G_{2}(\phi,X)~,~c=c_{(2)}=\frac{\dot{\phi}^{2}}{2}G_{2X}(\phi,X)~,~m_{2}^{4}=\frac{\dot{\phi}^{4}}{4}G_{2XX}(\phi,X)~,~m_{3}^{3}=m_{4}^{2}=\tilde{m}_{4}^{2}=0~.
\ee
According to Eq. (\ref{epsilon}), the slow-roll parameters can be reduced to
\be
\epsilon_{(2)}=\frac{3c_{(2)}}{c_{(2)}+\Lambda_{(2)}}=\frac{3\dot\phi^2G_{2X}}{2(\dot\phi^2G_{2X}-G_2)}~.
\ee
For canonical scalar field with $G_2(X,\phi)=X-V(\phi)$, the above definition of $\epsilon_{(2)}$ can be connected with the ratio between kinetic and potential terms $\gamma\equiv K/V$: $\epsilon_{(2)}=\gamma/(1+\gamma)$. Note that it is also coincide with another common-used definition: $\dot\phi^2G_{2X}/H^2$.

From Eqs. (\ref{c1})-(\ref{c4}), we have
\bea
&&D=4H^{2}M_{p}^{4}~,~c_{1}=M_{p}^{2}\left(\epsilon_{(2)} +2\frac{m_{2}^{4}}{H^{2}M_{p}^{2}}\right)~,~c_{2}=M_{p}^{2}~,~c_{3}=\frac{a}{H}M_{p}^{2}~,~c_{s}^{2}=\frac{H^{2}\epsilon_{(2)} M_{p}^{2}}{H^{2}\epsilon_{(2)} M_{p}^{2}+2m_{2}^{4}}~,\nonumber\\
&&{\cal D}_{T}=\frac{M_p^2}{8}~,~c_{T}^{2}=1~.
\eea
so using Eq. (\ref{r_final}), one has
\be
r=\frac{16\sqrt{2}HM_{p}\epsilon_{(2)}^{3/2}}{\sqrt{2H^{2}\epsilon_{(2)} M_{p}^{2}+\dot{\phi}^{4}G_{2XX}}}=\frac{16\epsilon_{(2)}^{3/2}}{\sqrt{\epsilon_{(2)}+2\delta_{KXX}}}~,
\ee
where we define the parameter $\delta_{KXX}\equiv\dot{\phi}^{4}G_{2XX}/(4H^2fM_p^2)$, with $f$ being unity in the current case.

One can see that for all the parameter choices of $\epsilon_{(2)}$ and $\delta_{KXX}$, the result recovers the standard consistency relation: $r=16\epsilon c_s$. Moreover, for the canonical scalar field $G_{2XX}=0$, one has $\delta_{KXX}=0$ and $c_s=1$, which is so simple that for given potential form, $\epsilon$ and $r$ can be directly related to parameters such as the power-law index of the potential, and the number of e-foldings \cite{Baumann:2009ds}. Therefore, in the canonical case, the only way of getting the small $r$ is to have small $\epsilon$ by requiring the potential to be very flat. Two specific examples are obvious in the literature: the small-field inflation with $\epsilon\sim constant$ (hilltop inflation \cite{Boubekeur:2005zm} for example) and the ultra-slow-roll inflation with $\epsilon\sim \eta^{-6}$, which is proposed by \cite{Kinney:2005vj} (with further studies in e.g., \cite{Namjoo:2012aa}) and extended to constant-roll inflation \cite{Motohashi:2014ppa}.

For noncanonical scalar field, the sound speed $c_s$ also plays its role. For $G_{2XX}\sim M_2^4>0$ which causes $c_s<1$, $r$ can be suppressed by $c_s$. Examples include ghost condensate inflation \cite{ArkaniHamed:2003uz} as well as DBI inflation \cite{Alishahiha:2004eh}. However, as for the single-field inflation, the non-Gassanianties is also related to $c_s$ in terms of $f_{nl}\sim c_s^{-2}$ roughly \cite{Chen:2006nt} where $f_{nl}$ describes the amplitude of non-Gaussianities. These models will also meet the danger of making $f_{nl}$ too large to be consistent with the current constraints \cite{Akrami:2019izv}.

\begin{figure}[ht]
\begin{center}
\includegraphics[width=8cm]{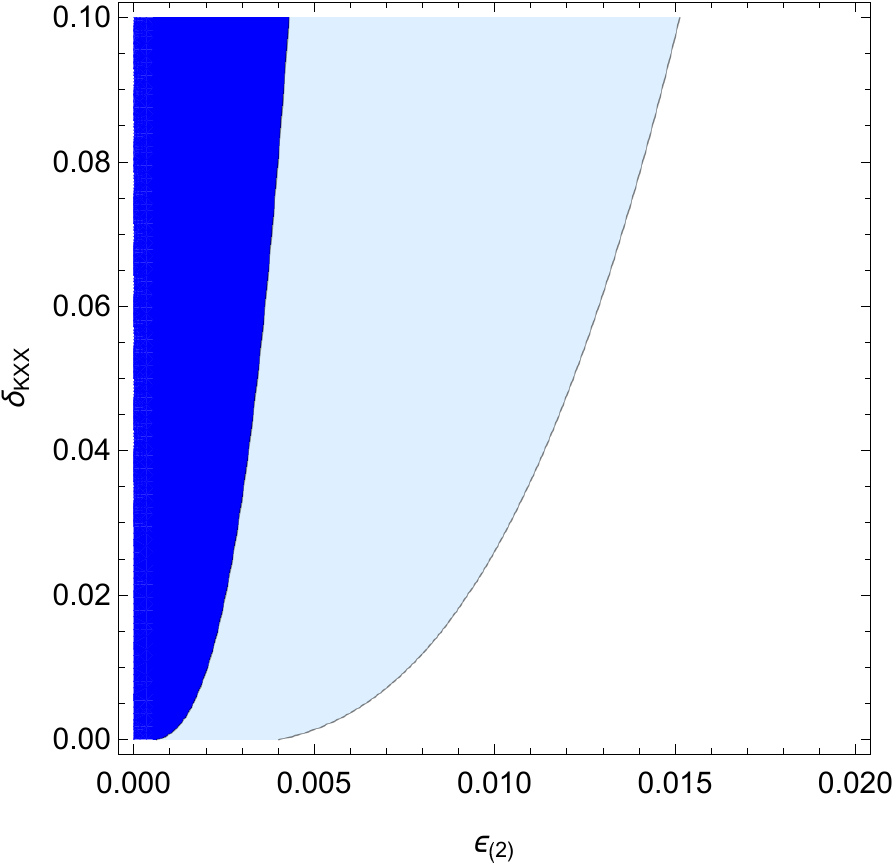}
\caption{The plot of $r$ in parameter space $[\epsilon_{(2)}, \delta_{KXX}]$. Light blue region denotes $0.01<r<0.064$, and blue region denotes $r<0.01$.} \label{plot1}
\end{center}
\end{figure}

The plot of $r$ in parameter space $[\epsilon_{(2)}, \delta_{KXX}]$ is shown in Fig. \ref{plot1}, where light blue region denotes $0.01<r<0.064$, and blue region denotes $r<0.01$. Note that when $\delta_{KXX}=0$, according to the consistency relation $r=16\epsilon$, $r<0.064$ and $r<0.001$ take $\epsilon$ into the very narrow region of $\epsilon<0.004$ and $\epsilon<0.000625$, respectively. However, when taking $\delta_{KXX}$ into consideration, the allowed region of $\epsilon_{(2)}$ will get much enlarged and larger $\epsilon_{(2)}$ will also be allowed. Moreover, considering the constraints of stability, namely $c_1>0$, $c_s^2>0$, we must restrict $\epsilon_{(2)}$ and $\delta_{KXX}$ to be within the region of $\epsilon_{(2)}>0$ and $\epsilon_{(2)}+2\delta_{KXX}>0$.

For more complicated case, however, not only the consistency relation $r=16\epsilon c_s$ will be violated, $\epsilon$ could also be affected and be deviated from $\epsilon_{(2)}$, due to the alterations of both background energy density and pressure, either of which will play a role of suppressing $r$. In order to illustrate, in the following, we write
\be
\Lambda=\Lambda_{(2)}+\Delta\Lambda~,~c=c_{(2)}+\Delta c~,
\ee
where $\Delta\Lambda$ and $\Delta c$ denotes the derivation of $\Lambda$ and $c$ from those in the case of K-essence. 
Thus we have
\be
\epsilon=\epsilon_{(2)}\left[1+\delta_{f}^{(1)}-\frac{1}{3}\frac{\Delta c}{M_{p}^{2}H^{2}f}-\frac{1}{3}\frac{\Delta\Lambda}{M_{p}^{2}H^{2}f}\right]+\frac{\Delta c}{M_{p}^{2}H^{2}f}+\frac{1}{2}\delta_{f}^{(1)}(\delta_{f}^{(2)}-1)
\ee
so the effects on $\epsilon$ can be corresponded to several parameters such as $\delta_{f}^{(1)}$, $\delta_{f}^{(2)}$, $\frac{\Delta c}{M_{p}^{2}H^{2}f}$ and $\frac{\Delta\Lambda}{M_{p}^{2}H^{2}f}$.

\subsection{Galileon: $M_{p}^{2}R/2+L_{2}+L_{3}$}
The next case to consider is the Galileon case, the proposal of which is inspired by the ghost problems that appeared in DGP models \cite{Nicolis:2008in}. Although the original proposal introduced the Galileon symmetry, it was later extended to general case that include also dependence of the field itself, with the addition of the term $G_3(\phi, X)\Box\phi$ \cite{Deffayet:2010qz}. When $G_3$ contains $\phi$ only, this term coincides with the kinetic term only by moduli a total derivative. Here for simplicity we take $G_3$ to be of the form $G_3=g(\phi)X$, therefore can be written in terms of action (\ref{eft_action}) with the correspondence:
\bea
&&f=1~,~\Lambda=\frac{\dot{\phi}^{2}}{2}G_{2X}(\phi,X)-G_{2}(\phi,X)-\frac{\dot{\phi}^{2}}{2}(\ddot{\phi}+3H\dot{\phi})g(\phi)~,~c=\frac{\dot{\phi}^{2}}{2}G_{2X}(\phi,X)+\frac{\dot{\phi}^{2}}{2}(\ddot{\phi}-3H\dot{\phi})g(\phi)+\frac{\dot{\phi}^{4}}{2}g_{\phi}(\phi)~,\nonumber\\
&&m_{2}^{4}=\frac{\dot{\phi}^{4}}{4}G_{2XX}(\phi,X)-\frac{\dot{\phi}^{2}}{4}(\ddot{\phi}+3H\dot{\phi})g(\phi)+\frac{\dot{\phi}^{4}}{4}g_{\phi}(\phi)~,~m_{3}^{3}=-\dot{\phi}^{3}g(\phi)~,~m_{4}^{2}=\tilde{m}_{4}^{2}=0~,
\eea
then from Eqs. (\ref{c1})-(\ref{c4}), we have
\bea
&&D=(2HM_{p}^{2}-m_{3}^{3})^{2}~,~c_{1}=\frac{M_{p}^{2}}{(2HM_{p}^{2}-m_{3}^{3})^{2}}(3m_{3}^{6}+4H^{2}\epsilon M_{p}^{4}+8m_{2}^{4}M_{p}^{2})~,~c_{2}=M_{p}^{2}~,\nonumber\\
&&c_{3}=\frac{2aM_{p}^{4}}{2HM_{p}^{2}-m_{3}^{3}}~,~c_{s}^{2}=\frac{4H^{2}\epsilon M_{p}^{4}+2HM_{p}^{2}m_{3}^{3}-m_{3}^{6}+2(m_{3}^{3})^{\cdot}M_{p}^{2}}{4H^{2}\epsilon M_{p}^{4}+8m_{2}^{4}M_{p}^{2}+3m_{3}^{6}}~,\nonumber\\
&&{\cal D}_{T}=\frac{M_p^2}{8}~,~c_{T}^{2}=1~.
\eea
so using Eq. (\ref{r_final}), one has
\be
r=\frac{16}{(2HM_{p}^{2}-m_{3}^{3})^{2}}\frac{\left[4H^{2}\epsilon M_{p}^{4}+2HM_{p}^{2}m_{3}^{3}-m_{3}^{6}+2(m_{3}^{3})^{\cdot}M_{p}^{2}\right]^{3/2}}{\left(4H^{2}\epsilon M_{p}^{4}+8m_{2}^{4}M_{p}^{2}+3m_{3}^{6}\right)^{1/2}}~.
\ee
Moreover, in order to take into account of the variation of $\epsilon$, we note that
\bea
\frac{\Delta c}{M_{p}^{2}H^{2}f}&=&\frac{g(\phi)\dot{\phi}^{3}}{2M_{p}^{2}H}(\frac{\ddot{\phi}}{H\dot{\phi}}-3)+\frac{g_{\phi}(\phi)\dot{\phi}^{4}}{2M_{p}^{2}H^{2}}~,\\
\frac{\Delta\Lambda}{M_{p}^{2}H^{2}f}&=&-\frac{g(\phi)\dot{\phi}^{3}}{2M_{p}^{2}H}(\frac{\ddot{\phi}}{H\dot{\phi}}+3)~,\\
\delta_{f}^{(1)}&=&\delta_{f}^{(2)}=0~.
\eea

It is useful to define the ``slow-varying" parameters as $\delta_{gX}\equiv\dot{\phi}^{3}g(\phi)/(HM_{p}^{2})$, $\delta_{g\phi}\equiv g_{\phi}(\phi)\dot{\phi}^{4}/(H^{2}M_{p}^{2})$, and $\delta_{\phi}\equiv\ddot{\phi}/H\dot{\phi}$. Therefore the above formula will become $\frac{\Delta c}{M_{p}^{2}H^{2}f}=\frac{1}{2}\delta_{gX}(\delta_{\phi}-3)+\frac{1}{2}\delta_{g\phi}$, $\frac{\Delta\Lambda}{M_{p}^{2}H^{2}f}=-\frac{1}{2}\delta_{gX}(\delta_{\phi}+3)$, thus
\be
\epsilon=\epsilon_{(2)}(1+\delta_{gX}-\frac{1}{6}\delta_{g\phi})+\frac{1}{2}\delta_{gX}(\delta_{\phi}-3)+\frac{1}{2}\delta_{g\phi}~.
\ee
Moreover, we have $m_{3}^{3}=-\dot{\phi}^{3}g(\phi)=-HM_{p}^{2}\delta_{gX}$, $(m_{3}^{3})^{\cdot}=-g_{\phi}(\phi)\dot{\phi}^{4}-3g(\phi)\dot{\phi}^{2}\ddot{\phi}=-H^{2}M_{p}^{2}\delta_{g\phi}-3H^{2}M_{p}^{2}\delta_{gX}\delta_{\phi}$, and $m_{2}^{4}=H^{2}M_{p}^{2}\delta_{KXX}-\frac{1}{4}H^{2}M_{p}^{2}\delta_{gX}(\delta_{\phi}+3)+\frac{1}{4}H^{2}M_{p}^{2}\delta_{g\phi}$ where $\delta_{KXX}$ has already been defined previously. In this regard, we get
\be
\label{r_case3}
r=\frac{16}{(2+\delta_{gX})^{2}}\frac{\left[4\epsilon_{(2)}+4\epsilon_{(2)}\delta_{gX}-2\epsilon_{(2)}\delta_{g\phi}/3-8\delta_{gX}-4\delta_{gX}\delta_{\phi}-\delta_{gX}^{2}\right]^{3/2}}{\left(4\epsilon_{(2)}+4\epsilon_{(2)}\delta_{gX}-2\epsilon_{(2)}\delta_{g\phi}/3-12\delta_{gX}+4\delta_{g\phi}+3\delta_{gX}^{2}+8\delta_{KXX}\right)^{1/2}}~.
\ee

Since one can notice that the dependence of $r$ on $\delta_{KXX}$ is also obvious and is the same as in the Kessence case, we will also ignore it in the following discussions. We now consider a simple case in which $g_{\phi}(\phi)\simeq0$ ($g\simeq const.$), namely $\delta_{g\phi}\simeq0$, therefore $\epsilon$ and $r$ will become
\bea
\epsilon&=&\epsilon_{(2)}(1+\delta_{gX})+\frac{1}{2}\delta_{gX}(\delta_{\phi}-3)~,\\
r&=&\frac{16}{(2+\delta_{gX})^{2}(s+1)^{3}}\frac{\left[4\epsilon_{(2)}+4\epsilon_{(2)}\delta_{gX}-8\delta_{gX}-4\delta_{gX}\delta_{\phi}-\delta_{gX}^{2}\right]^{3/2}}{\left(4\epsilon_{(2)}+4\epsilon_{(2)}\delta_{gX}-12\delta_{gX}+3\delta_{gX}^{2}\right)^{1/2}}~.
\eea
Furtherly consider all the parameters are smaller than 1, then the terms of parameters multiplied or squared can be viewed as higher order infinitesimals. Therefore $r$ could be greatly simplified as:
 \be
 r\simeq 16 \frac{(\epsilon_{(2)}-2\delta_{gX})^{3/2}}{(\epsilon_{(2)}-3\delta_{gX})^{1/2}}~.
 \ee
 For usual case of $\epsilon_{(2)}>0$, whether $\delta_{gX}$ is positive or negative, the inclusion of $\delta_{gX}$ will cause the raise of $r$, so we tend to get large $r$ as in G-inflation model \cite{Kobayashi:2010cm}. However, for $\epsilon_{(2)}<0$, which makes phantom-like inflation \cite{Piao:2004tq} possible to happen, there is certain room for small $r$ where $\delta_{gX}$ is also negative. Note also that due to the requirement of stabilities ($c_1>0$, $c_s^2>0$), the case of positive $\delta_{gX}$ for $\epsilon_{(2)}<0$ is forbidden.

\begin{figure}[ht]
\begin{center}
\includegraphics[width=18cm]{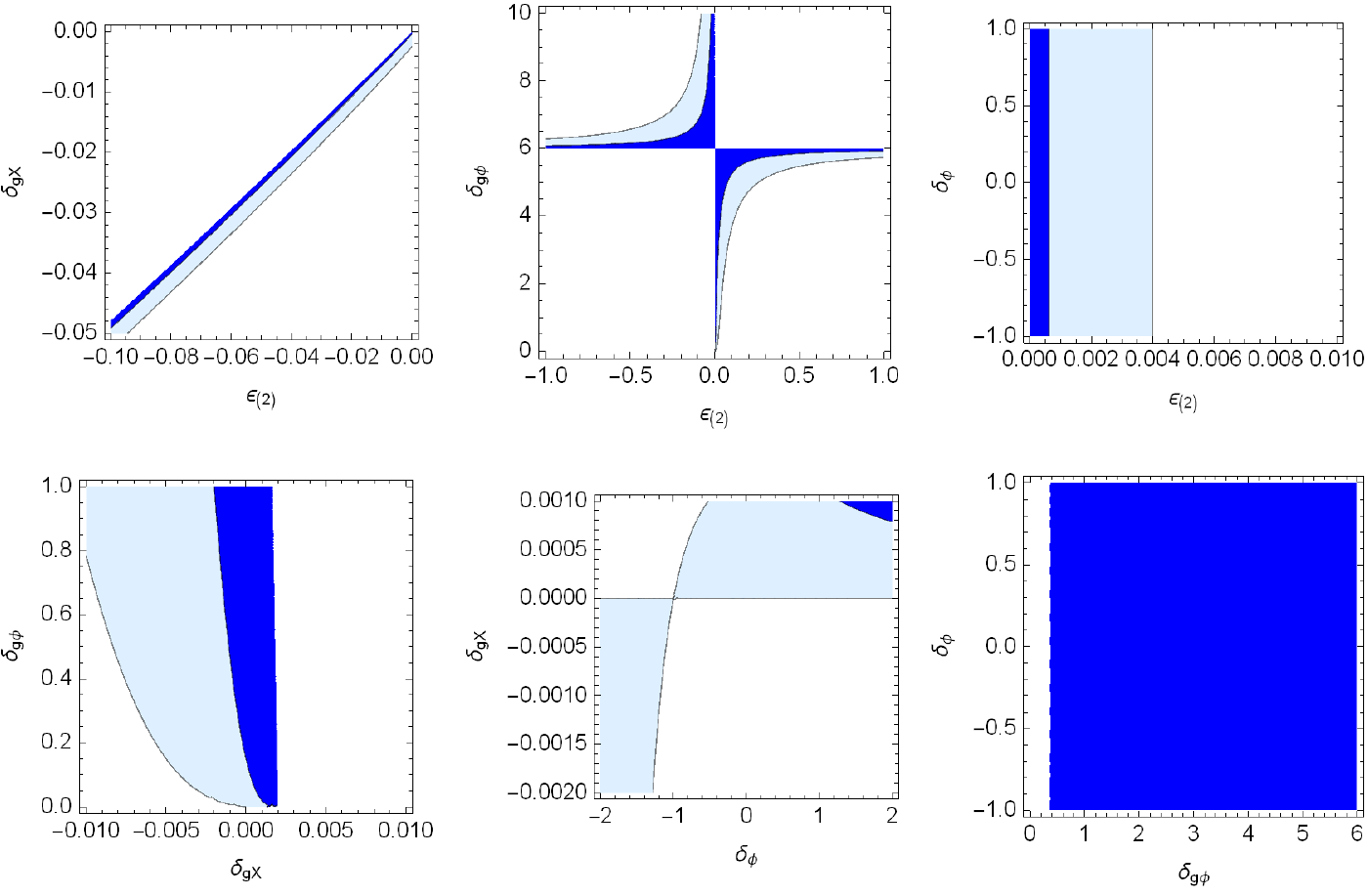}
\caption{The plot of $r$ in parameter space $[\epsilon_{(2)}, \delta_{gX}]$ (setting $\delta_{g\phi}=0$, $\delta_{\phi}=0$), $[\epsilon_{(2)}, \delta_{g\phi}]$ (setting $\delta_{gX}=0$, $\delta_{\phi}=0$), $[\epsilon_{(2)}, \delta_{\phi}]$ (setting $\delta_{gX}=0$, $\delta_{g\phi}=0$), $[\delta_{gX}, \delta_{g\phi}]$ (setting $\epsilon_{(2)}=0.004$, $\delta_{\phi}=0$), $[\delta_{gX}, \delta_{\phi}]$ (setting $\epsilon_{(2)}=0.004$, $\delta_{g\phi}=0$), $[\delta_{g\phi}, \delta_{\phi}]$ (setting $\epsilon_{(2)}=0.004$, $\delta_{gX}=0$). Light blue region denotes $0.01<r<0.064$, and blue region denotes $r<0.01$.} \label{plot2}
\end{center}
\end{figure}

\begin{figure}[ht]
\begin{center}
\includegraphics[width=18cm]{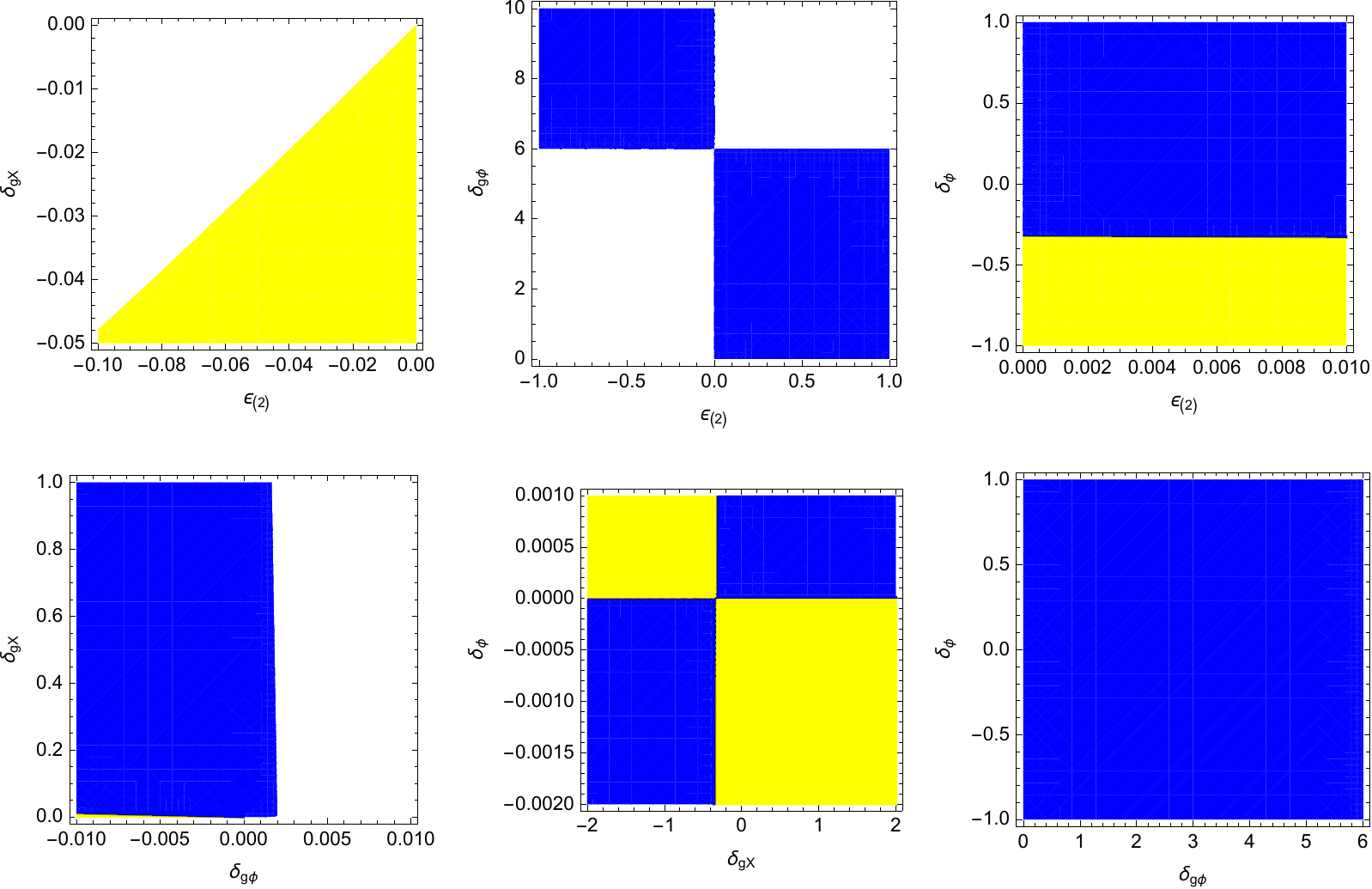}
\caption{The plot of $r-16\epsilon c_s$ in parameter space $[\epsilon_{(2)}, \delta_{gX}]$ (setting $\delta_{g\phi}=0$, $\delta_{\phi}=0$), $[\epsilon_{(2)}, \delta_{g\phi}]$ (setting $\delta_{gX}=0$, $\delta_{\phi}=0$), $[\epsilon_{(2)}, \delta_{\phi}]$ (setting $\delta_{gX}=0$, $\delta_{g\phi}=0$), $[\delta_{gX}, \delta_{g\phi}]$ (setting $\epsilon_{(2)}=0.004$, $\delta_{\phi}=0$), $[\delta_{gX}, \delta_{\phi}]$ (setting $\epsilon_{(2)}=0.004$, $\delta_{g\phi}=0$), $[\delta_{g\phi}, \delta_{\phi}]$ (setting $\epsilon_{(2)}=0.004$, $\delta_{gX}=0$). Blue region denotes $r-16\epsilon c_s<0$, and yellow region denotes $r-16\epsilon c_s>0$.} \label{plot2C}
\end{center}
\end{figure}

The plot of $r$ in parameter space $[\epsilon_{(2)}, \delta_{gX}, \delta_{g\phi}, \delta_{\phi}]$ is shown in Fig. \ref{plot2}, where light blue region denotes $0.01<r<0.064$, and blue region denotes $r<0.01$. According to the plot, several remarks are as follows: 1) In the first plot, the allowed region extend to where $\epsilon_{(2)}<0$, making possible for phantom inflation; 2) For other cases, there are also large space for $r$ to be within (0.064, 0.1); 3) Note that for $\delta_{gX}=0$, $r$ can be small for $\delta_{g\phi}\simeq 6$, which is because there is a pole in the numerator of $r$ in Eq. (\ref{r_case3}). However it seems difficult to make consistency of both $\delta_{gX}=0$ and nonvanishing $\delta_{g\phi}$; 4) From the rightmost column one can see that for $\delta_{gX}=0$, the dependence of $\delta_{\phi}$ is also decoupled. In the allowed region for $\epsilon_{(2)}$ and $\delta_{g\phi}$, the room for small $r$ is also quite large.

The ratio between $r$ and $16\epsilon c_{s}$ can also be straightforwardly calculated as:
\be
\frac{r}{16\epsilon c_{s}}=\frac{\left[4\epsilon_{(2)}+4\epsilon_{(2)}\delta_{gX}-2\epsilon_{(2)}\delta_{g\phi}/3-8\delta_{gX}-4\delta_{gX}\delta_{\phi}-\delta_{gX}^{2}\right]}{(2+\delta_{gX})^{2}\left[\epsilon_{(2)}(1+\delta_{gX}-\frac{1}{6}\delta_{g\phi})+\frac{1}{2}\delta_{gX}(\delta_{\phi}-3)+\frac{1}{2}\delta_{g\phi}\right]}~,
\ee
which depends on all the parameters, and we also plot this comparison in Fig. \ref{plot2C} (in order to avoid the possible pole in the dominator, we instead plot $r-16\epsilon c_{s}$ and compare it with $0$.), with blue color for $r<16\epsilon c_s$, while yellow color for $r>16\epsilon c_s$. Therefore for Galileon field, the consistency relationship is usually broken.

\subsection{Modified gravity: $L_{2}+L_{4}$}
Here we consider the case where $G_4$ terms involves in, which includes nonminimally coupling case as well as $f(R)$ modified gravity case. For simplicity, we take $G_4$ to be purely function of $\phi$, namely $G_{4}=M_{p}^{2}A(\phi)/2$ where $A(\phi)$ is an arbitrary function. Therefore this case can be written in terms of action (\ref{eft_action}) with the correspondence:
\be
f=A(\phi)~,~\Lambda=\frac{\dot{\phi}^{2}}{2}G_{2X}(\phi,X)-G_{2}(\phi,X)~,~c=\frac{\dot{\phi}^{2}}{2}G_{2X}(\phi,X)~,~m_{2}^{4}=\frac{\dot{\phi}^{4}}{4}G_{2XX}(\phi,X)~,~m_{3}^{3}=m_{4}^{2}=\tilde{m}_{4}^{2}=0~.
\ee
From Eqs. (\ref{c1})-(\ref{c4}), we have
\bea
&&D=(2AH+\dot{A})^{2}M_{p}^{4}~,~c_{1}	=\frac{1}{(2AH+\dot{A})^{2}}A(4A^{2}H^{2}\epsilon M_{p}^{2}+3\dot{A}^{2}M_{p}^{2}+8AM_{2}^{4}-2A\ddot{A}M_{p}^{2}+2A\dot{A}HM_{p}^{2})~,\nonumber\\
&&c_{2}=AM_{p}^{2}~,~c_{3}=\frac{2aA^{2}M_{p}^{2}}{2AH+\dot{A}}~,~c_{s}^{2}=\frac{2A\dot{A}H+3\dot{A}^{2}+4A^{2}H^{2}\epsilon-2A\ddot{A}}{2A\dot{A}H+3\dot{A}^{2}+4A^{2}H^{2}\epsilon-2A\ddot{A}+8A\frac{m_{2}^{4}}{M_{p}^{2}}}~,\nonumber\\
&&{\cal D}_T=\frac{M_p^2A(\phi)}{8}~,~c_{T}=1~.
\eea

Defining $\delta_{A}^{(1)}\equiv\dot{A}/HA$, $\delta_{A}^{(2)}\equiv\ddot{A}/H\dot{A}$, and using Eq. (\ref{r_final}), we get
\be
r=\frac{16}{(2+\delta_{A}^{(1)})^{2}}\frac{(2\delta_{A}^{(1)}+3\delta_{A}^{(1)2}+4\epsilon-2\delta_{A}^{(1)}\delta_{A}^{(2)})^{3/2}}{(2\delta_{A}^{(1)}+3\delta_{A}^{(1)2}+4\epsilon-2\delta_{A}^{(1)}\delta_{A}^{(2)}+8\delta_{KXX})^{1/2}}~.
\ee
Moreover, in this case we have $\frac{\Delta c}{M_{p}^{2}H^{2}f}=0$, $\frac{\Delta\Lambda}{M_{p}^{2}H^{2}f}=0$, $\delta_{f}^{(1)}=\delta_{A}^{(1)}$, $\delta_{f}^{(2)}=\delta_{A}^{(2)}$, therefore
\be
\epsilon=\epsilon_{(2)}(1+\delta_{A}^{(1)})+\frac{1}{2}\delta_{A}^{(1)}(\delta_{A}^{(2)}-1)
\ee
so
\bea
r&=&\frac{16}{(2+\delta_{A}^{(1)})^{2}}\frac{[+3\delta_{A}^{(1)2}+4\epsilon_{(2)}(1+\delta_{A}^{(1)})]^{3/2}}{[+3\delta_{A}^{(1)2}+4\epsilon_{(2)}(1+\delta_{A}^{(1)})+8\delta_{KXX}]^{1/2}}~,\nonumber\\
c_{s}^{2}&=&\frac{2\delta_{A}^{(1)}+3\delta_{A}^{(1)2}+4\epsilon-2\delta_{A}^{(1)}\delta_{A}^{(2)}}{2\delta_{A}^{(1)}+3\delta_{A}^{(1)2}+4\epsilon-2\delta_{A}^{(1)}\delta_{A}^{(2)}+8\delta_{KXX}]^{1/2}}~.
\eea

For the same reason as the above case, we ignore the effects of $\delta_{KXX}$. In this case, $c_s^2$ will exactly be unity, and one could get a very neat form of $r$:
\be
r\simeq\frac{16(3\delta_{A}^{(1)2}+4\epsilon_{(2)}+4\epsilon_{(2)}\delta_{A}^{(1)})}{(2+\delta_{A}^{(1)})^{2}}
\ee
which only involves two parameters. Since now $c_s=1$, one can see that $r$ has obviously deviated from the consistency relation: $r=16\epsilon c_s$. Moreover, we also find that $c_T=1$, namely even gravity is modified in the current form, the sound speed of tensor perturbation is unaltered. This indicates that $c_T$ can only deviate from unity in a more complicated form of modified gravity, e.g., when there is a kinetic coupling to the gravity, as will be demonstrated in the next case. The interest in the deviation of $c_T$ from unity is spurred by the constraints imposed by the latest GW event from a binary neutron star merger, namely GW170817, and its electromagnetic counterpart GRB170817A, see \cite{Monitor:2017mdv}.

Let us now turn to another interesting case, namely where $G_2(\phi, X)=G_2(\phi)$ is a pure function of $\phi$. This, by conformal transformation, is nothing but $f(R)$ gravity \cite{DeFelice:2010aj} (see also \cite{Nojiri:2006ri}). In this case we have $\epsilon=\frac{1}{2}\delta_{A}^{(1)}\delta_{A}^{(2)}-\frac{1}{2}\delta_{A}^{(1)}$, therefore
\bea
r&=&\frac{16(2\delta_{A}^{(1)}+3\delta_{A}^{(1)2}+4\epsilon-2\delta_{A}^{(1)}\delta_{A}^{(2)})}{(2+\delta_{A}^{(1)})^{2}}~\nonumber\\
&=&\frac{48\delta_{A}^{(1)2}}{(2+\delta_{A}^{(1)})^{2}}~.
\eea
namely the number of parameters involved again get reduced, up to only one. Moreover, since $r$ is proportional to $\delta_{A}^{(1)2}$, it is easier to get small $r$, as long as $\delta_{A}^{(1)}$ is not too large. For example, for $r<0.064(0.01)$, one needs $-0.070(-0.028)<\delta_A^{(1)}<0.076(0.029)$. A concrete example is the famous inflation model proposed by Starobinsky \cite{Starobinsky:1980te}.

\begin{figure}[ht]
\begin{center}
\includegraphics[width=16cm]{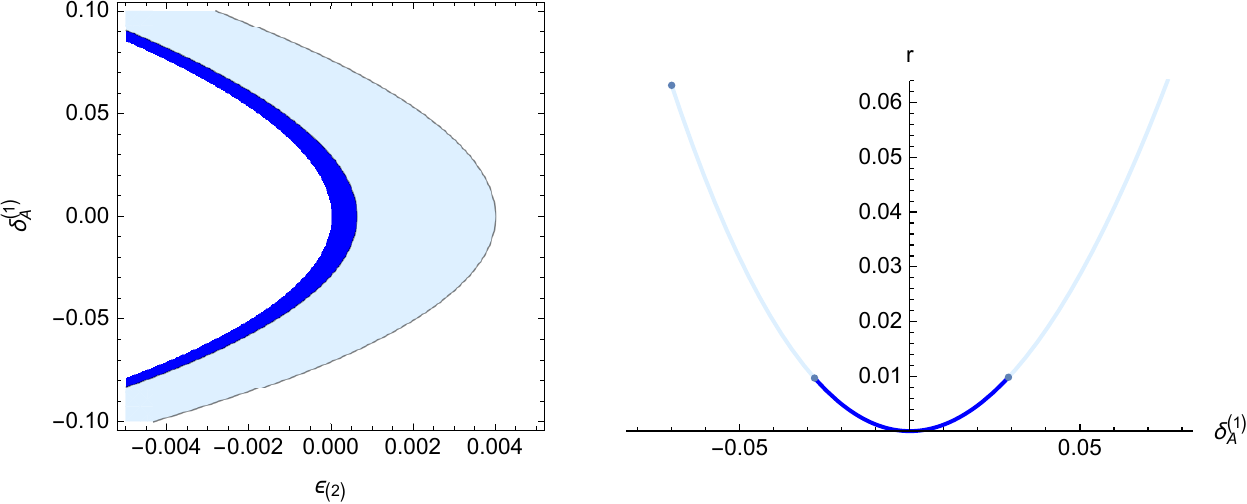}
\caption{The plot of $r$ in parameter space ${\epsilon_{(2)}, \delta_{A}^{(1)}}$ for nonminimal coupling models (Left panel), and vs. $\delta_{A}^{(1)}$ for $f(R)$ modified gravity models. Light blue region denotes $0.01<r<0.064$, and blue region denotes $r<0.01$.} \label{plot3}
\end{center}
\end{figure}

\begin{figure}[ht]
\begin{center}
\includegraphics[width=16cm]{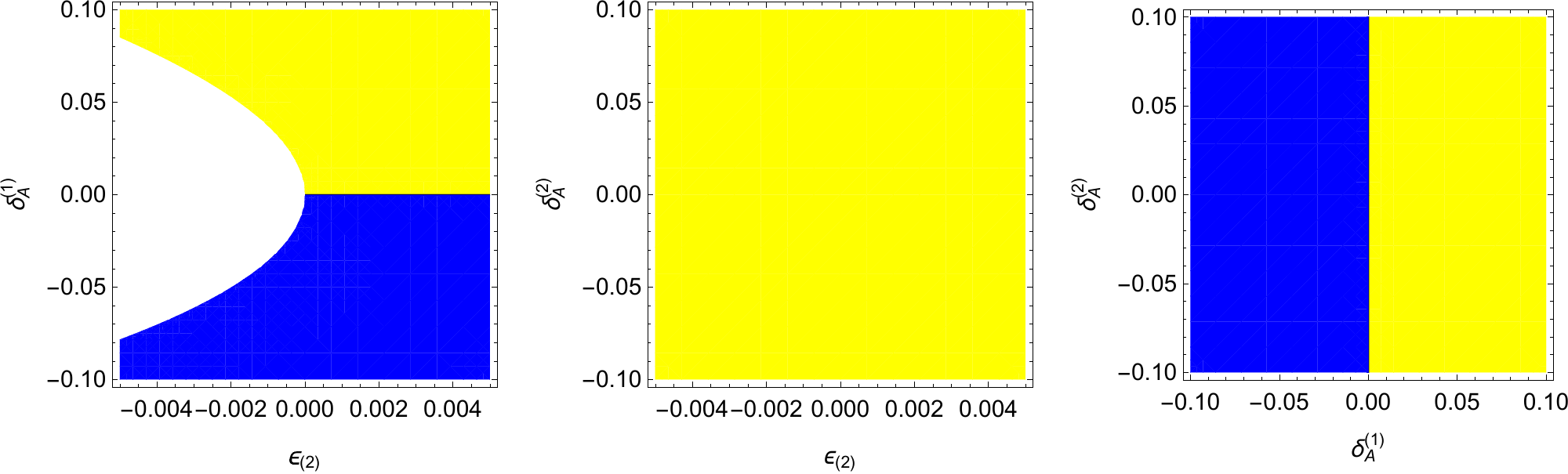}
\caption{The plot of $r-16\epsilon c_s$ in parameter space $[\epsilon_{(2)}, \delta_{A}^{(1)}]$ (setting $\delta_{A}^{(2)}=0$), $[\epsilon_{(2)}, \delta_{A}^{(2)}]$ (setting $\delta_{A}^{(1)}=0.01$), $[\delta_{A}^{(1)}, \delta_{A}^{(2)}]$ (setting $\epsilon_{(2)}=0.004$). Blue region denotes $r-16\epsilon c_s<0$, and yellow region denotes $r-16\epsilon c_s>0$. } \label{plot3C}
\end{center}
\end{figure}

The plot of $r$ in parameter space $[\epsilon_{(2)}, \delta_{A}^{(1)}]$ is shown in Fig. \ref{plot3}, where light blue region denotes $0.01<r<0.064$, and blue region denotes $r<0.01$. Note that the allowed region extend to where $\epsilon_{(2)}<0$, which means that in presence of nonminimal coupling, $r$ can be within the detectable region even for phantom inflation, but in this case, large $\delta_{A}^{(1)}$ is needed in order not to cause the instabilities. For $f(R)$ gravity, there is modest space for $\delta_{A}^{(1)}$ to make $r$ within the detectable region. The comparison of $r$ to the consistency relation $r=16\epsilon c_s$ is also shown in Fig. \ref{plot3C}. One can see that for the trivial case of $\delta_{A}^{(1)}=0$, one can calculate that $r=16\epsilon c_s$ and consistency relation is recovered. However, for positive/negative $\delta_{A}^{(1)}$, one has $r$ larger/smaller than $16\epsilon c_s$.

\subsection{Nonminimal derivative coupling: $M_{p}^{2}R/2+L_{2}+L_{5}$}
In last case we discussed about nonminimal coupling of field and gravity but missed the possibility of derivative coupling, namely the gravity coupling with derivatives of the inflaton field. This is a very different case from above, which will probably even involve $G_5$. Here we assume $G_5=h(\phi)$ is a pure function of $\phi$, which is equivalent to the introduction of a term like $G_{\mu\nu}\partial^\mu\phi\partial^\nu\phi$, up to a total derivative \cite{Amendola:1993uh}. This case can be written in terms of action (\ref{eft_action}) with the correspondence:
\bea
&&f=1-h_{\phi}(\phi)\dot{\phi}^{2}~,\\
&&\Lambda=\frac{\dot{\phi}^{2}}{2}G_{2X}(\phi,X)-G_{2}(\phi,X)-\frac{1}{2}M_{p}^{2}H\dot{f}-2M_{p}^{2}(f-1)\dot{H}-\frac{1}{2}M_{p}^{2}\ddot{f}+3M_{p}^{2}H^{2}(f-1)~,\\
&&c=\frac{\dot{\phi}^{2}}{2}G_{2X}(\phi,X)+\frac{7}{2}M_{p}^{2}H\dot{f}+2M_{p}^{2}(f-1)\dot{H}+\frac{1}{2}M_{p}^{2}\ddot{f}+9M_{p}^{2}H^{2}(f-1)~,\\
&&m_{2}^{4}=\frac{\dot{\phi}^{4}}{4}G_{2XX}(\phi,X)+\frac{1}{4}M_{p}^{2}H\dot{f}+M_{p}^{2}(f-1)\dot{H}+\frac{1}{4}M_{p}^{2}\ddot{f}~,\\
&&m_{3}^{3}=M_{p}^{2}\dot{f}+4M_{p}^{2}H(f-1)~,~m_{4}^{2}=-M_{p}^{2}(f-1)~,~\tilde{m}_{4}^{2}=-M_{p}^{2}(f-1)~.
\eea
From Eqs. (\ref{c1})-(\ref{c4}), we have
\bea
D&=&4H^{2}M_{p}^{4}(3f-4)^{2}=4H^{2}M_{p}^{4}(1+3h_{\phi}\dot{\phi}^{2})^{2}~,\\
c_{1}&=&\frac{(2-f)M_{p}^{2}}{4H^{2}(3f-4)^{2}}\{48H^{2}(f-1)^{2}+(2-f)[2\dot{\phi}^{4}G_{2XX}(\phi,X)/M_{p}^{2}+4H\dot{f}+4H^{2}\epsilon(2-f)]\}~,\\
c_{2}&=&fM_{p}^{2}=(1-h_{\phi}\dot{\phi}^{2})M_{p}^{2}~,\\
c_{3}	&=&\frac{-aM_{p}^{2}(2-f)^{2}}{H(3f-4)}~,\\
c_{s}^{2}&=&\frac{4H^{2}}{(2-f)}\frac{\{\dot{f}(3f-2)(2-f)/H-(2-f)^{2}(3f-4)(1+\epsilon)-f(3f-4)^{2}\}}{\{48H^{2}(f-1)^{2}+(2-f)[2\dot{\phi}^{4}G_{2XX}(\phi,X)/M_{p}^{2}+4H\dot{f}+4H^{2}\epsilon(2-f)]\}}~,\\	
{\cal D}_T&=&\frac{M_p^2}{8}(2-f)=\frac{M_p^2}{8}(1+h_{\phi}\dot{\phi}^{2})~,\\
c_{T}	&=&\frac{f}{Q_{T}}=\frac{f}{2-f}=\frac{1-h_{\phi}\dot{\phi}^{2}}{1+h_{\phi}\dot{\phi}^{2}}~,
\eea
so using Eq. (\ref{r_final}), one has
\be
r=\frac{32}{(3f-4)^{2}}\frac{[2\dot{f}(2-f)/H-(2-f)^{2}(3f-4)(1+\epsilon)-f(3f-4)^{2}]^{3/2}}{\{48(f-1)^{2}(2-f)+(2-f)^{2}[2\dot{\phi}^{4}G_{2XX}(\phi,X)/(H^2M_{p}^{2})+4\dot{f}/H+4\epsilon(2-f)]\}^{1/2}f^{3/2}}
\ee		

For ease of our analysis, we define $\delta_{h\phi}\equiv f_{(5)}/f$, where $f_{(5)}=f-1$. Therefore we have $\dot{f_{(5)}}=\dot{f}$, $\ddot{f_{(5)}}=\ddot{f}$, and $\dot{\delta}_{h\phi}=(f_{(5)}/f)^{\cdot}=\dot{f}_{(5)}/f-f_{(5)}\dot{f}/f^{2}=(\dot{f}/f)(1-f_{(5)}/f)=H\delta_{f}^{(1)}(1-\delta_{h\phi})$. Moreover, to avoid ghost and gradient instabilities in this model (${\cal D}_T>0$, $c_T>0$), we require $-1<h_\phi\dot\phi^2<1$, so from the definition one finds $\delta_{h\phi}<1/2$ needed even for non-inflationary models.

Moreover, in this case we have
\bea
\frac{\Delta c}{M_{p}^{2}H^{2}f}&=&-\frac{M_{p}^{2}H\dot{f}_{(5)}}{2M_{p}^{2}H^{2}f}-\frac{2M_{p}^{2}f_{(5)}\dot{H}}{M_{p}^{2}H^{2}f}-\frac{M_{p}^{2}\ddot{f}_{(5)}}{2M_{p}^{2}H^{2}f}+\frac{3M_{p}^{2}H^{2}f_{(5)}}{M_{p}^{2}H^{2}f}\\
&=&-\frac{1}{2}\delta_{f}^{(1)}+2\epsilon\delta_{h\phi}-\frac{1}{2}\delta_{f}^{(1)}\delta_{f}^{(2)}+3\delta_{h\phi}~,\\
\frac{\Delta\Lambda}{M_{p}^{2}H^{2}f}&=&+\frac{7M_{p}^{2}H\dot{f}_{(5)}}{2M_{p}^{2}H^{2}f}+\frac{2M_{p}^{2}f_{(5)}\dot{H}}{M_{p}^{2}H^{2}f}+\frac{M_{p}^{2}\ddot{f}_{(5)}}{2M_{p}^{2}H^{2}f}+\frac{9M_{p}^{2}H^{2}f_{(5)}}{M_{p}^{2}H^{2}f}\\
&=&+\frac{7}{2}\delta_{f}^{(1)}-2\epsilon\delta_{h\phi}+\frac{1}{2}\delta_{f}^{(1)}\delta_{f}^{(2)}+9\delta_{h\phi}~,
\eea
therefore
\be
\epsilon=\frac{\epsilon_{(2)}(1-4\delta_{h\phi})-\delta_{f}^{(1)}+3\delta_{h\phi}}{1-2\delta_{h\phi}}~,
\ee
and
\bea
r=\frac{16}{(4\delta_{h\phi}-1)^{2}}\frac{[6\delta_{f}^{(1)}\delta_{h\phi}(1-2\delta_{h\phi})+\delta_{h\phi}(3-2\delta_{h\phi})(1-4\delta_{h\phi})+(1-4\delta_{h\phi})^{2}(1-2\delta_{h\phi})\epsilon_{(2)}]^{3/2}}{[3\delta_{h\phi}(1+2\delta_{h\phi})+(1-2\delta_{h\phi})(1-4\delta_{h\phi})\epsilon_{(2)}+2\delta_{KXX}(1-2\delta_{h\phi})]^{1/2}}~.
\eea

\begin{figure}[ht]
\begin{center}
\includegraphics[width=16cm]{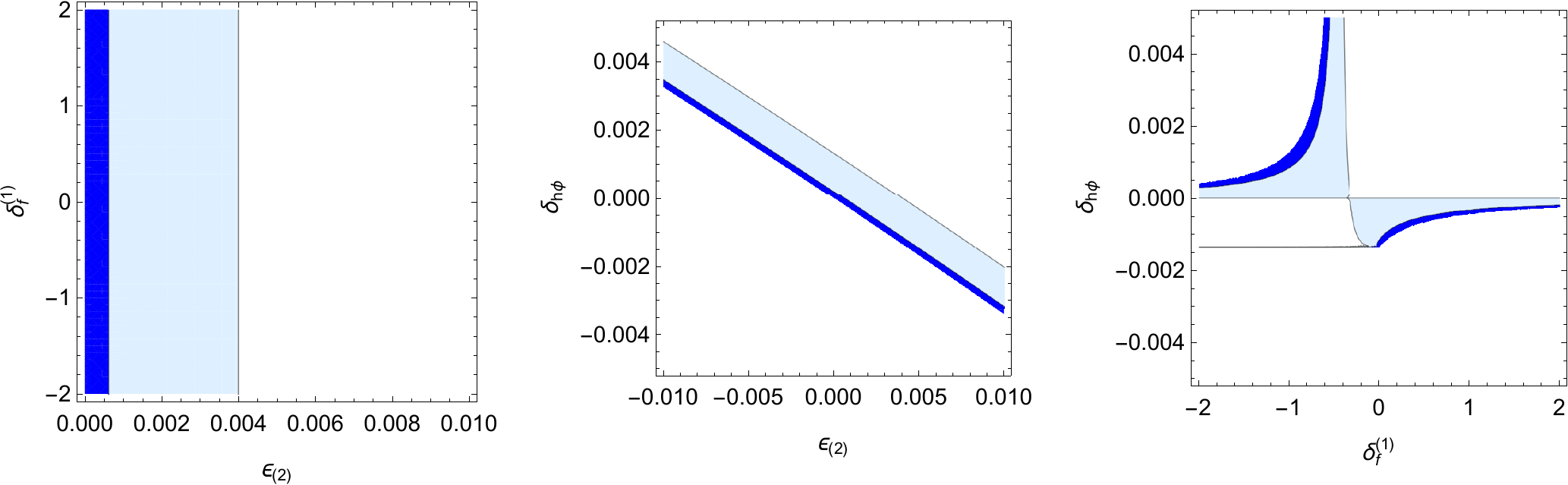}
\caption{The plot of $r$ in parameter space $[\epsilon_{(2)}, \delta_f^{(1)}]$ (setting $\delta_{h\phi}=0$), $[\epsilon_{(2)}, \delta_{h\phi}]$ (setting $\delta_f^{(1)}=0$), $[\delta_f^{(1)}, \delta_{h\phi}]$ (setting $\epsilon_{(2)}=0.004$). Light blue region denotes $0.01<r<0.064$, and blue region denotes $r<0.01$.} \label{plot4}
\end{center}
\end{figure}

\begin{figure}[ht]
\begin{center}
\includegraphics[width=16cm]{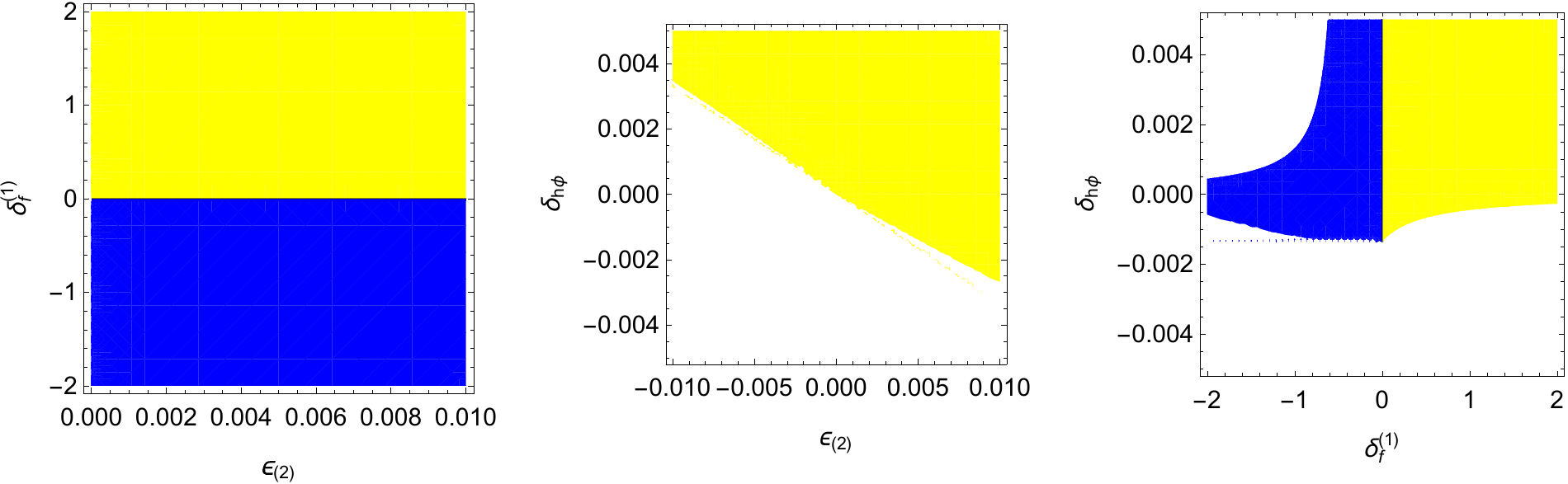}
\caption{The plot of $r-16\epsilon c_s$ in parameter space $[\epsilon_{(2)}, \delta_f^{(1)}]$ (setting $\delta_{h\phi}=0$), $[\epsilon_{(2)}, \delta_{h\phi}]$ (setting $\delta_f^{(1)}=0$), $[\delta_f^{(1)}, \delta_{h\phi}]$ (setting $\epsilon_{(2)}=0.004$). Blue region denotes $r-16\epsilon c_s<0$, and yellow region denotes $r-16\epsilon c_s>0$.} \label{plot4C}
\end{center}
\end{figure}

We now consider a simple case, namely $\delta_{f}^{(1)}\simeq0$, which means that $\delta_{h\phi}=const.$ Moreover, we also ignore the effect from $\delta_{KXX}$. Therefore $\epsilon$ and $r$ will be reduced to:
\bea
\epsilon&=&\frac{\epsilon_{(2)}(1-4\delta_{h\phi})+3\delta_{h\phi}}{1-2\delta_{h\phi}}~,\\
r&=&\frac{16[(\epsilon_{(2)}(1-4\delta_{h\phi})+3\delta_{h\phi})(1-2\delta_{h\phi})+4\delta_{h\phi}^{2}]^{3/2}}{[12\delta_{h\phi}^{2}(1-4\delta_{h\phi})+(\epsilon_{(2)}(1-4\delta_{h\phi})+3\delta_{h\phi})(1-2\delta_{h\phi})(1-4\delta_{h\phi})]^{1/2}}\nonumber\\
&\simeq&16[\epsilon_{(2)}+3\delta_{h\phi}+\mathcal{O}(\delta_{h\phi}^{2})]~,
\eea
where in the last step we expand the expression of $r$ in terms of slow-varying parameters. One could see that the $\epsilon_{(2)}$ and $\delta_{h\phi}$ is anti-proportional to each other in determining $r$.
Moreover, for $\delta_{h\phi}\simeq0$, $\delta_{f}^{(1)}\simeq0$ case ($f\simeq1$), we have
\be
\epsilon\simeq\epsilon_{(2)}~,~r=16\epsilon_{(2)}~\ee
which recovers the canonical field case.

The plot of $r$ in parameter space $[\epsilon_{(2)}, \delta_f^{(1)}, \delta_{h\phi}]$ is shown in Fig. \ref{plot4}, where light blue region denotes $0.01<r<0.064$, and blue region denotes $r<0.01$. One can see that (from middle panel) for $\delta_f^{(1)}=0$, the allowed region lies in where $\epsilon_{(2)}$ and $\delta_{h\phi}$ goes roughly anti-proportional to each other, and large $\epsilon_{(2)}$ is allowable for large negative $\delta_{h\phi}$. Like nonminimal coupling case, negative $\epsilon_{(2)}$ (phantom inflation) is also allowable for large positive $\delta_{h\phi}$. Moreover, when $\delta_{h\phi}=0$, the dependence of $r$ on $\delta_f^{(1)}$ is also decoupled (left panel). The comparison of $r$ to the consistency relation $r=16\epsilon c_s$ is also shown in Fig. \ref{plot4C}. From the left panel we can see that, for $\delta_{h\phi}=0$, the deviation from the consistency relation $r=16\epsilon c_s$ (in this case $c_s=1$) totally depends on $\delta_f^{(1)}$, namely $r>/<16\epsilon$ when $\delta_f^{(1)}>/<0$, this is because in this case,  $r$ go back to the trivial case $r=16\epsilon_{(2)}=16(\epsilon+\delta_f^{(1)})$.

\subsection{Beyond Horndeski theory: $M_{p}^{2}R/2+L_{2}+L_{6}$}
The last case we consider is the beyond Horndeski model. 
In \cite{Cai:2016thi, Cai:2017tku}, 
we considered the beyond Horndeski model, which can realize a non-singular universe without either ghost or gradient instabilities, 
by introducing a nonzero $\Delta\tilde{m}_{4}^{2}$ in the action (\ref{eft_action}). 
Here we assume that $\tilde{m}_{4}^{2}=\Delta\tilde{m}_{4}^{2}=q(\phi)M_p^2$ being a pure function of $\phi$ and $q(\phi)$ is a dimensionless function, then this case can be written in terms of action (\ref{eft_action}) with the correspondence:
\bea
&&f=1~,~\Lambda=\frac{\dot{\phi}^{2}}{2}G_{2X}(\phi,X)-G_{2}(\phi,X)~,~c=\frac{\dot{\phi}^{2}}{2}G_{2X}(\phi,X)~,\nonumber\\
&&m_{2}^{4}=\frac{\dot{\phi}^{4}}{4}G_{2XX}(\phi,X)~,~m_{3}^{3}=m_{4}^{2}=0~,~\tilde{m}_{4}^{2}=q(\phi)M_p^2~.
\eea
From Eqs. (\ref{c1})-(\ref{c4}), we have
\bea
&&D=4H^2M_{p}^{4}~,~c_{1}=\frac{1}{H^{2}}(H^{2}\epsilon M_{p}^{2}+2m_{2}^{4})~,~c_{2}=M_{p}^{2}~,\\
&&c_{3}=\frac{aM_{p}^{2}}{H}\left(1+2\frac{\tilde{m}_{4}^{2}}{M_{p}^{2}}\right)~,~c_{s}^{2}=\frac{H^{2}\epsilon M_{p}^{2}+2H^{2}\tilde{m}_{4}^{2}(1+\epsilon)-2H\dot{\tilde{m}}_{4}^{2}}{H^{2}\epsilon M_{p}^{2}+2m_{2}^{4}}~,\\		
&&{\cal D}_T=\frac{M_p^2}{8}~,~c_{T}=1~.
\eea

In this case, we have $\frac{\Delta c}{M_{p}^{2}H^{2}f}=\frac{\Delta c}{M_{p}^{2}H^{2}f}=\delta_f^{(1)}=\delta_f^{(2)}=0$, therefore $\epsilon=\epsilon_{(2)}$, namely there is no derivation of $\epsilon$ from the effect of $L_6$. Define $\delta_q=\dot q/(Hq)$, we have
\bea
r=\frac{16[\epsilon_{(2)} +2q(1+\epsilon_{(2)}-\delta_{q})]^{3/2}}{(\epsilon_{(2)}+8\delta_{KXX})^{1/2}}~,
\eea

From the expression we can see that, for the case of $\delta_q\simeq 0$ ($q(\phi)$ is nearly constant), to have smaller $r$ than that is given by consistency relation, one must have $q<0$ ($\delta_{KXX}$ also ignored). However, the constraint from positivity of $c_1$ and $c_s^2$ requires that $\epsilon_{(2)}+8\delta_{KXX}>0$ as well as $\epsilon_{(2)}+2q(1+\epsilon_{(2)})-\delta_q>0$.

\begin{figure}[ht]
\begin{center}
\includegraphics[width=16cm]{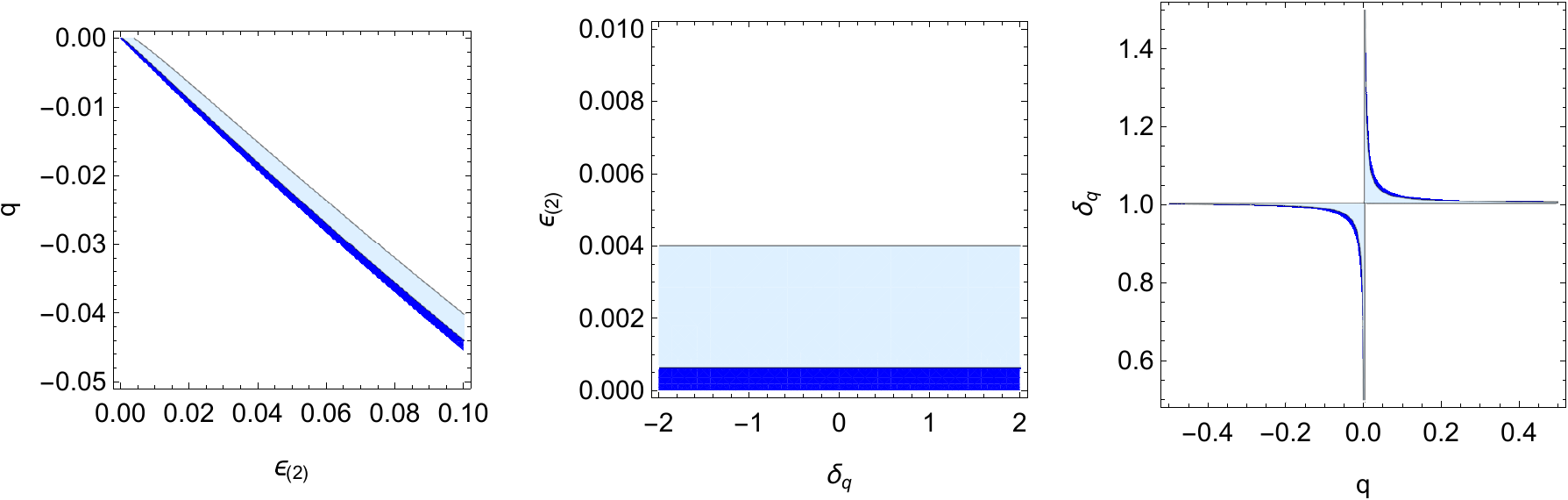}
\caption{The plot of $r$ in parameter space $[\epsilon_{(2)}, q]$ (setting $\delta_{q}=0$), $[\epsilon_{(2)}, \delta_{q}]$ (setting $q=0$), $[q, \delta_{q}]$ (setting $\epsilon_{(2)}=0.004$). Light blue region denotes $0.01<r<0.064$, and blue region denotes $r<0.01$.} \label{plot5}
\end{center}
\end{figure}

\begin{figure}[ht]
\begin{center}
\includegraphics[width=16cm]{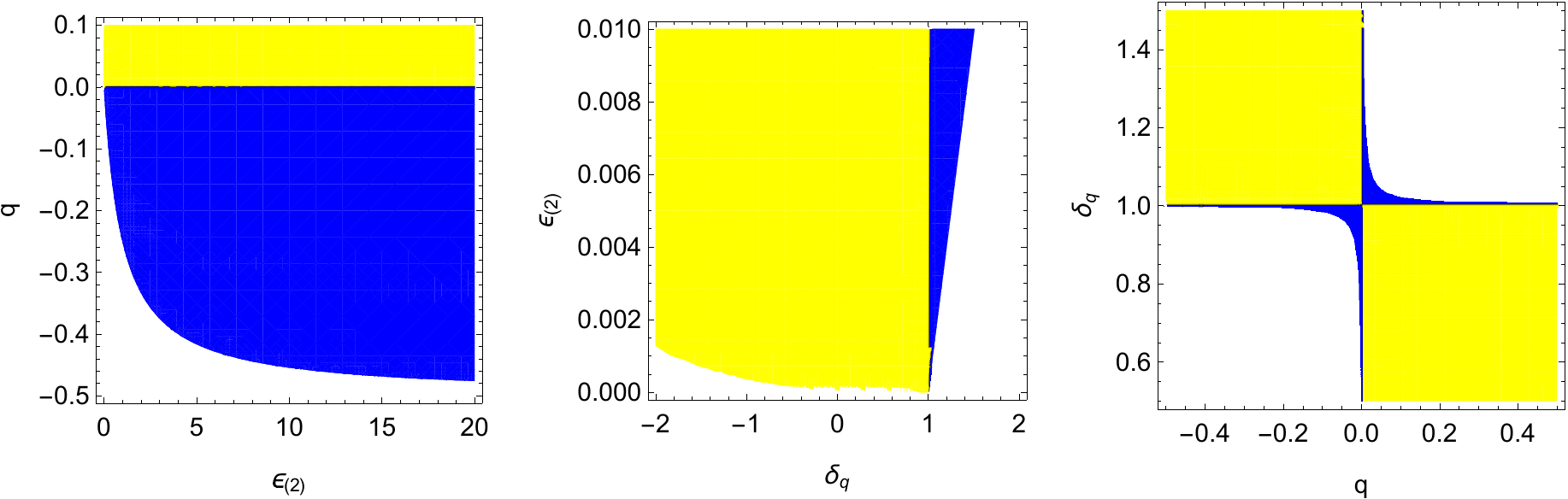}
\caption{The plot of $r-16\epsilon c_s$ in parameter space $[\epsilon_{(2)}, q]$ (setting $\delta_{q}=0$), $[\epsilon_{(2)}, \delta_{q}]$ (setting $q=0$), $[q, \delta_{q}]$ (setting $\epsilon_{(2)}=0.004$). Blue region denotes $r-16\epsilon c_s<0$, and yellow region denotes $r-16\epsilon c_s>0$.} \label{plot5C}
\end{center}
\end{figure}

The plot of $r$ in parameter space $[\epsilon_{(2)}, q, \delta_{q}]$ is shown in Fig. \ref{plot5}, where light blue region denotes $0.01<r<0.064$, and blue region denotes $r<0.01$. For $q<0$, small $r$ can still be obtained (left panel). For $q=0$, $r$ go back to the trivial case $r=16\epsilon_{(2)}$ and the dependence of $r$ on $\delta_{q}$ is also decoupled (middle panel). For $q$ around $0$ and $\delta_{q}$ around $1$, there still exists room for small $r$ even when $\epsilon_{(2)}=0.004$, which is at the edge of $r=0.064$ in the absense of $L_6$ (right panel). The comparison of $r$ to the consistency relation $r=16\epsilon c_s$ is also shown in Fig. \ref{plot5C}.

\section{conclusion}
The next decade will be a Gravitational Wave decade, with many more experiments of GWs getting down to work, and many more signals of GWs will be discovered. Especially, the ambitious ground-based experiment AliCPT in Tibet, China, is aiming to search for signals of PGWs with improved accuracy in the coming few years \cite{Li:2017lat}. The current and future constraints has divided the amplitude of $r$ into three parts, namely $r>0.064$ (disfavored by current data), $r\in (0.064,0.01)$ (within the observable window of next experiments like AliCPT) as well as $r<0.01$ (still waiting for further detections). In this paper, we have formulated the tensor/scalar ratio in the generic setup with the facility of the EFT approach. As an application, we theoretically studied which kind of inflation models let $r$ fall into the last two regions. In each model, we have analyzed the relation between $r$ and other slow-varying parameters and obtained the corresponding regions in parameter spaces. Furthermore, we have also discussed the deviation of $r$ from the consistency relation in each model. 
We summarize our conclusive remarks in the following:

1. Making use of the EFT approach, the tensor/scalar ratio $r$ for the given action (\ref{eft_action}) can be expressed as in Eq. (\ref{r_final}). Note that this expression is applicable for both cases where the power spectrum is constant or growing for scalar and tensor perturbations.

2. From the expression, one can see the running of sound speed affects $r$ in an obvious way. When $|s+1|>1$ or $|s_T+1|<1$, $r$ will get suppressed, or vise versa, where we simply assumed $c_s\sim \eta^s$, $c_T\sim \eta^{s_T}$. Note that the conclusion might not be applicable to more complicated cases.

3. For Kessence model where only $M_p^2R/2$ and $L_2$ are involved in the Lagrangian, $r$ is in accord with the consistency relation, $r=16\epsilon c_s$. In this case, various ways can be done to suppress $r$, e.g., to have small $\epsilon$, or small sound speed, by non-canonicity of the scalar field. However, the small sound speed will meet the danger of large non-Gaussianities.

4. For Lagrangians beyond $L_2$, $r$ will deviate from the consistency relation, which will cause another mechanism for small $r$. Moreover, this allows for phantom inflation to give rise to small $r$.

5. For nonminimal coupling theory, namely $L_4$ with $G_4=G_4(\phi)$, the sound speed of tensor perturbations is still unity, $c_T=1$. However, it will change when kinetic coupling to gravity also involves. Phantom inflation is allowed to have a small $r$ in these cases.

6. Small $r$ can also be obtained by taking into account the beyond-Horndeski part, namely $L_6$, even in the absence of $L_3$, $L_4$, and $L_5$.

Other than getting results of each concrete model in the ordinary model-by-model analysis, 
we have confirmed them at a more general level with the help of the EFT approach. 
By showing more detailed and more precise relationships between $r$ and those parameters, our analysis and numerical plots will be useful for concrete model-buildings. 
For instance, we can use the present method to analyze other subclasses of the Horndeski theory, or beyond Horndeski theories. 
We can also enlarge our current scope, for example, by furtherly turning on the parameters $\bar{m}_4$, $\bar{m}_5$, $\bar{\lambda}$ and $\tilde{\lambda}$, to include models with higher-order spatial derivatives. 

Before ending, let us remind that for the current discussions, we have focused only on $r$ and have not taken into account constraints from other variables on the early universe models, such as spectral index (and even its running) and non-Gaussianities. Although we assumed that the power spectra of these models are scale-invariant, it deserves to consider how such a scale-invariance would impose constraints on those model parameters; and with the observational data being more and more precise, the running of the index will also become another important constraining tool. Moreover, the non-Gaussianities is also an interesting probe of the early Universe. In Ref. \cite{Quintin:2015rta}, authors discussed that in matter bounce scenario driven by Horndeski theory, one could not get small $r$ while keeping $f_{nl}$ small enough to be within the current constraints (a.k.a. no-go theorem), while it is also interesting to consider such constraints for other early universe scenarios/models (examples has been given in \cite{Akama:2019qeh}). We will address the above discussion in future works.

\begin{acknowledgments}
We thank Pierre Zhang, Yun-Song Piao and Masahide Yamaguchi for helpful discussions. We also acknowledge Jun Chen for his early contributions to this work. This work was supported by the National Natural Science Foundation of China with Grants No.~11653002 and No.~11875141.

\end{acknowledgments}


\begin{thebibliography}{99}

\bibitem{Abbott:2017xzu}
  B.~P.~Abbott {\it et al.} [LIGO Scientific and Virgo and 1M2H and Dark Energy Camera GW-E and DES and DLT40 and Las Cumbres Observatory and VINROUGE and MASTER Collaborations],
  Nature {\bf 551}, no. 7678, 85 (2017)
  [arXiv:1710.05835 [astro-ph.CO]].

\bibitem{Abbott:2016blz}
  B.~P.~Abbott {\it et al.} [LIGO Scientific and Virgo Collaborations],
  Phys.\ Rev.\ Lett.\  {\bf 116}, no. 6, 061102 (2016)
  [arXiv:1602.03837 [gr-qc]].

\bibitem{Abbott:2016nmj}
  B.~P.~Abbott {\it et al.} [LIGO Scientific and Virgo Collaborations],
  Phys.\ Rev.\ Lett.\  {\bf 116}, no. 24, 241103 (2016)
  [arXiv:1606.04855 [gr-qc]].

\bibitem{Abbott:2017vtc}
  B.~P.~Abbott {\it et al.} [LIGO Scientific and VIRGO Collaborations],
  Phys.\ Rev.\ Lett.\  {\bf 118}, no. 22, 221101 (2017)
  Erratum: [Phys.\ Rev.\ Lett.\  {\bf 121}, no. 12, 129901 (2018)]
  [arXiv:1706.01812 [gr-qc]].

\bibitem{Abbott:2017gyy}
  B.~. P.~.Abbott {\it et al.} [LIGO Scientific and Virgo Collaborations],
  Astrophys.\ J.\  {\bf 851}, no. 2, L35 (2017)
  [arXiv:1711.05578 [astro-ph.HE]].

\bibitem{Abbott:2017oio}
  B.~P.~Abbott {\it et al.} [LIGO Scientific and Virgo Collaborations],
  Phys.\ Rev.\ Lett.\  {\bf 119}, no. 14, 141101 (2017)
  [arXiv:1709.09660 [gr-qc]].

\bibitem{TheLIGOScientific:2017qsa}
  B.~P.~Abbott {\it et al.} [LIGO Scientific and Virgo Collaborations],
  Phys.\ Rev.\ Lett.\  {\bf 119}, no. 16, 161101 (2017)
  [arXiv:1710.05832 [gr-qc]].

\bibitem{Nan:2011um}
  R.~Nan {\it et al.},
  Int.\ J.\ Mod.\ Phys.\ D {\bf 20}, 989 (2011)
  [arXiv:1105.3794 [astro-ph.IM]].

\bibitem{Somiya:2011np}
  K.~Somiya [KAGRA Collaboration],
  Class.\ Quant.\ Grav.\  {\bf 29}, 124007 (2012)
  [arXiv:1111.7185 [gr-qc]].

\bibitem{AmaroSeoane:2012km}
  P.~Amaro-Seoane {\it et al.},
  GW Notes {\bf 6}, 4 (2013)
  [arXiv:1201.3621 [astro-ph.CO]].

\bibitem{Luo:2015ght}
  J.~Luo {\it et al.} [TianQin Collaboration],
  Class.\ Quant.\ Grav.\  {\bf 33}, no. 3, 035010 (2016)
  [arXiv:1512.02076 [astro-ph.IM]].

\bibitem{Guo:2018npi}
  Z.~K.~Guo, R.~G.~Cai and Y.~Z.~Zhang,
  arXiv:1807.09495 [gr-qc].

\bibitem{Ferdman:2010xq}
  R.~D.~Ferdman {\it et al.},
  Class.\ Quant.\ Grav.\  {\bf 27}, 084014 (2010)
  [arXiv:1003.3405 [astro-ph.HE]].

\bibitem{Li:2017lat}
  Y.~P.~Li, Y.~Liu, S.~Y.~Li, H.~Li and X.~Zhang,
  arXiv:1709.09053 [astro-ph.IM];
  H.~Li {\it et al.},
  Published on line by National Science Review, 2018
  [arXiv:1710.03047 [astro-ph.CO]];
  H.~Li, S.~Y.~Li, Y.~Liu, Y.~P.~Li and X.~Zhang,
  Nat.\ Astron.\  {\bf 2}, no. 2, 104 (2018)
  [arXiv:1802.08455 [astro-ph.IM]].


\bibitem{Starobinsky:1979ty}
A.~A.~Starobinsky,
JETP Lett. \textbf{30}, 682-685 (1979)


\bibitem{Stewart:1993bc}
  E.~D.~Stewart and D.~H.~Lyth,
  Phys.\ Lett.\ B {\bf 302}, 171 (1993)
  [gr-qc/9302019].

\bibitem{Hu:1997hv}
  W.~Hu and M.~J.~White,
  New Astron.\  {\bf 2}, 323 (1997)
  [astro-ph/9706147].

\bibitem{Seljak:1996gy}
  U.~Seljak and M.~Zaldarriaga,
  Phys.\ Rev.\ Lett.\  {\bf 78}, 2054 (1997)
  [astro-ph/9609169];
  M.~Zaldarriaga and U.~Seljak,
  Phys.\ Rev.\ D {\bf 55}, 1830 (1997)
  [astro-ph/9609170].

\bibitem{Kamionkowski:1996zd}
  M.~Kamionkowski, A.~Kosowsky and A.~Stebbins,
  Phys.\ Rev.\ Lett.\  {\bf 78}, 2058 (1997)
  [astro-ph/9609132];
  M.~Kamionkowski, A.~Kosowsky and A.~Stebbins,
  Phys.\ Rev.\ D {\bf 55}, 7368 (1997)
  [astro-ph/9611125].

\bibitem{WMAP}
See the WMAP homepage, URL: https://map.gsfc.nasa.gov.

\bibitem{PLANCK}
See the PLANCK homepage, URL: http://www.esa.int/Our$\_$Activities/Space$\_$Science/Planck/.

\bibitem{Akrami:2018odb}
  Y.~Akrami {\it et al.} [Planck Collaboration],
  arXiv:1807.06211 [astro-ph.CO].

\bibitem{ACT}
See the ACT homepage, URL: https://act.princeton.edu.

\bibitem{POLARBEAR}
See the POLARBEAR homepage, URL: http://bolo.berkeley.edu/polarbear/.

\bibitem{SPT}
See the SPT homepage, URL: https://pole.uchicago.edu.

\bibitem{BICEP}
See the BICEP/Keck Array homepage, URL: https://www.cfa.harvard.edu/CMB/keckarray/.

\bibitem{Starobinsky:1980te}
  A.~A.~Starobinsky,
  Phys.\ Lett.\ B {\bf 91}, 99 (1980)
  [Phys.\ Lett.\  {\bf 91B}, 99 (1980)]
  [Adv.\ Ser.\ Astrophys.\ Cosmol.\  {\bf 3}, 130 (1987)].

\bibitem{Kinney:2005vj}
  W.~H.~Kinney,
  Phys.\ Rev.\ D {\bf 72}, 023515 (2005)
  [gr-qc/0503017].

\bibitem{Motohashi:2014ppa}
  H.~Motohashi, A.~A.~Starobinsky and J.~Yokoyama,
  JCAP {\bf 1509}, 018 (2015)
  [arXiv:1411.5021 [astro-ph.CO]];
  Z.~Yi and Y.~Gong,
  JCAP {\bf 1803}, 052 (2018)
  [arXiv:1712.07478 [gr-qc]].



\bibitem{Cheung:2007st}
  C.~Cheung, P.~Creminelli, A.~L.~Fitzpatrick, J.~Kaplan and L.~Senatore,
  JHEP {\bf 0803}, 014 (2008)
  [arXiv:0709.0293 [hep-th]].

\bibitem{Gubitosi:2012hu}
  G.~Gubitosi, F.~Piazza and F.~Vernizzi,
  JCAP {\bf 1302}, 032 (2013)
  [JCAP {\bf 1302}, 032 (2013)]
  [arXiv:1210.0201 [hep-th]].

\bibitem{Gleyzes:2013ooa}
  J.~Gleyzes, D.~Langlois, F.~Piazza and F.~Vernizzi,
  JCAP {\bf 1308}, 025 (2013)
  [arXiv:1304.4840 [hep-th]].

\bibitem{Piazza:2013coa}
  F.~Piazza and F.~Vernizzi,
  Class.\ Quant.\ Grav.\  {\bf 30}, 214007 (2013)
  [arXiv:1307.4350 [hep-th]].

\bibitem{Kase:2014cwa}
  R.~Kase and S.~Tsujikawa,
  Int.\ J.\ Mod.\ Phys.\ D {\bf 23}, no. 13, 1443008 (2015)
  [arXiv:1409.1984 [hep-th]].

\bibitem{Gao:2014soa}
  X.~Gao,
  Phys.\ Rev.\ D {\bf 90}, 081501 (2014)
  [arXiv:1406.0822 [gr-qc]];
  X.~Gao,
  Phys.\ Rev.\ D {\bf 90}, 104033 (2014)
  [arXiv:1409.6708 [gr-qc]].

\bibitem{Cai:2016thi}
  Y.~Cai, Y.~Wan, H.~G.~Li, T.~Qiu and Y.~S.~Piao,
  JHEP {\bf 1701}, 090 (2017)
  [arXiv:1610.03400 [gr-qc]].

\bibitem{Cai:2017tku}
  Y.~Cai, H.~G.~Li, T.~Qiu and Y.~S.~Piao,
  Eur.\ Phys.\ J.\ C {\bf 77}, no. 6, 369 (2017)
  [arXiv:1701.04330 [gr-qc]].
  
\bibitem{Horndeski:1974wa}
  G.~W.~Horndeski,
  Int.\ J.\ Theor.\ Phys.\  {\bf 10}, 363 (1974).
  For review, see
  T.~Kobayashi,
  Rept. Prog. Phys. \textbf{82} (2019) no.8, 086901
  [arXiv:1901.07183 [gr-qc]].

  
\bibitem{Gleyzes:2014dya}
  J.~Gleyzes, D.~Langlois, F.~Piazza and F.~Vernizzi,
  Phys. Rev. Lett. \textbf{114} (2015) no.21, 211101
  [arXiv:1404.6495 [hep-th]].
  
\bibitem{DeFelice:2014bma}
  A.~De Felice and S.~Tsujikawa,
  Phys.\ Rev.\ D \textbf{91} (2015) no.10, 103506
  [arXiv:1411.0736 [hep-th]].

\bibitem{Deffayet:2009mn}
  C.~Deffayet, S.~Deser and G.~Esposito-Farese,
  Phys.\ Rev.\ D {\bf 80}, 064015 (2009)
  [arXiv:0906.1967 [gr-qc]];
  C.~Deffayet, X.~Gao, D.~A.~Steer and G.~Zahariade,
  Phys.\ Rev.\ D {\bf 84}, 064039 (2011)
  [arXiv:1103.3260 [hep-th]].
  
\bibitem{ArmendarizPicon:1999rj}
  C.~Armendariz-Picon, T.~Damour and V.~F.~Mukhanov,
  Phys.\ Lett.\ B {\bf 458}, 209 (1999)
  [hep-th/9904075];
  C.~Armendariz-Picon, V.~F.~Mukhanov and P.~J.~Steinhardt,
  Phys.\ Rev.\ D {\bf 63}, 103510 (2001)
  [astro-ph/0006373].

\bibitem{Baumann:2009ds}
  D.~Baumann,
  arXiv:0907.5424 [hep-th].

\bibitem{Boubekeur:2005zm}
  L.~Boubekeur and D.~H.~Lyth,
  JCAP {\bf 0507}, 010 (2005)
  [hep-ph/0502047].

\bibitem{Namjoo:2012aa}
  M.~H.~Namjoo, H.~Firouzjahi and M.~Sasaki,
  EPL {\bf 101}, no. 3, 39001 (2013)
  [arXiv:1210.3692 [astro-ph.CO]];
  J.~Martin, H.~Motohashi and T.~Suyama,
  Phys.\ Rev.\ D {\bf 87}, no. 2, 023514 (2013)
  [arXiv:1211.0083 [astro-ph.CO]].

\bibitem{ArkaniHamed:2003uz}
  N.~Arkani-Hamed, P.~Creminelli, S.~Mukohyama and M.~Zaldarriaga,
  JCAP {\bf 0404}, 001 (2004)
  [hep-th/0312100].

\bibitem{Alishahiha:2004eh}
  M.~Alishahiha, E.~Silverstein and D.~Tong,
  Phys.\ Rev.\ D {\bf 70}, 123505 (2004)
  [hep-th/0404084].

\bibitem{Chen:2006nt}
  X.~Chen, M.~x.~Huang, S.~Kachru and G.~Shiu,
  JCAP {\bf 0701}, 002 (2007)
  [hep-th/0605045].

\bibitem{Akrami:2019izv}
  Y.~Akrami {\it et al.} [Planck Collaboration],
  arXiv:1905.05697 [astro-ph.CO].

\bibitem{Nicolis:2008in}
  A.~Nicolis, R.~Rattazzi and E.~Trincherini,
  Phys.\ Rev.\ D {\bf 79}, 064036 (2009)
  [arXiv:0811.2197 [hep-th]];
  C.~Deffayet, G.~Esposito-Farese and A.~Vikman,
  Phys.\ Rev.\ D {\bf 79}, 084003 (2009)
  [arXiv:0901.1314 [hep-th]];
  A.~Nicolis, R.~Rattazzi and E.~Trincherini,
  JHEP {\bf 1005}, 095 (2010)
  [Erratum-ibid.\  {\bf 1111}, 128 (2011)]
  [arXiv:0912.4258 [hep-th]].

\bibitem{Deffayet:2010qz}
  C.~Deffayet, O.~Pujolas, I.~Sawicki and A.~Vikman,
  JCAP {\bf 1010}, 026 (2010)
  [arXiv:1008.0048 [hep-th]];
  A.~De Felice and S.~Tsujikawa,
  Phys.\ Rev.\ D {\bf 84}, 124029 (2011)
  [arXiv:1008.4236 [hep-th]].

\bibitem{Kobayashi:2010cm}
  T.~Kobayashi, M.~Yamaguchi and J.~Yokoyama,
  Phys.\ Rev.\ Lett.\  {\bf 105}, 231302 (2010)
  [arXiv:1008.0603 [hep-th]].

\bibitem{Piao:2004tq}
  Y.~S.~Piao and Y.~Z.~Zhang,
  Phys.\ Rev.\ D {\bf 70}, 063513 (2004)
  [astro-ph/0401231].

\bibitem{Monitor:2017mdv}
  B.~P.~Abbott {\it et al.} [LIGO Scientific and Virgo and Fermi-GBM and INTEGRAL Collaborations],
  Astrophys.\ J.\  {\bf 848}, no. 2, L13 (2017)
  [arXiv:1710.05834 [astro-ph.HE]].

\bibitem{DeFelice:2010aj}
  A.~De Felice and S.~Tsujikawa,
  Living Rev.\ Rel.\  {\bf 13}, 3 (2010)
  [arXiv:1002.4928 [gr-qc]].
  
\bibitem{Nojiri:2006ri}
  S.~Nojiri and S.~D.~Odintsov,
  eConf \textbf{C0602061} (2006), 06
  [arXiv:hep-th/0601213 [hep-th]].  .

\bibitem{Amendola:1993uh}
  L.~Amendola,
  Phys.\ Lett.\ B {\bf 301}, 175 (1993)
  [gr-qc/9302010];
  S.~Capozziello and G.~Lambiase,
  Gen.\ Rel.\ Grav.\  {\bf 31}, 1005 (1999)
  [gr-qc/9901051];
  S.~Capozziello, G.~Lambiase and H.~J.~Schmidt,
  Annalen Phys.\  {\bf 9}, 39 (2000)
  [gr-qc/9906051];
  N.~Yang, Q.~Fei, Q.~Gao and Y.~Gong,
  Class.\ Quant.\ Grav.\  {\bf 33}, no. 20, 205001 (2016)
  [arXiv:1504.05839 [gr-qc]].

\bibitem{Quintin:2015rta}
  J.~Quintin, Z.~Sherkatghanad, Y.~F.~Cai and R.~H.~Brandenberger,
  Phys.\ Rev.\ D {\bf 92}, no. 6, 063532 (2015)
  [arXiv:1508.04141 [hep-th]];
  Y.~B.~Li, J.~Quintin, D.~G.~Wang and Y.~F.~Cai,
  JCAP {\bf 1703}, 031 (2017)
  [arXiv:1612.02036 [hep-th]].

\bibitem{Akama:2019qeh}
  S.~Akama, S.~Hirano and T.~Kobayashi,
  Phys.\ Rev.\ D {\bf 101}, no. 4, 043529 (2020)
  [arXiv:1908.10663 [gr-qc]].
\end{thebibliography}
\end{document}